\documentclass[12pt,osajnl2,preprint,showpacs]{revtex4} 
\usepackage[draft]{hyperref} 
\DeclareRobustCommand{\baselinestretch{2.2}}

\usepackage{graphicx,natbib}

\newtheorem{theorem}{Theorem}

\newenvironment{proof}{~ \\[0.1in] {\bf Proof.} }{\hfill  \bigskip \\[0.1in]}

\begin{document}

\title{Propagation of Aberrations through Phase Induced Amplitude Apodization coronagraph}

\author{Laurent Pueyo,$^{1,2,*}$ N. Jeremy Kasdin,$^3$ and Stuart Shaklan$^{4}$}

\address{
$^1$NASA Sagan Fellow, Johns Hopkins University, Department of physics and astronomy, \\ 366 Bloomberg Center
3400 N. Charles Street, Baltimore, MD 21218 USA\\
$^2$ Space Telescope Science Institute, \\
 3700 San Marin Drive, Baltimore, MD 21218, USA\\
$^3$ Department of Mechanical and Aerospace Engineering, \\
Princeton University, Princeton, NJ, 08544, USA \\
$^4$ Jet Propulsion Laboratory, California Insitute of Technology, \\
4800 Oak Grove Drive, Pasadena, CA, 91109, USA\\
$^*$Corresponding author: lap@pha.jhu.edu
}
%
%
%
%
%


\begin{abstract}
The specification of polishing requirements for the optics in coronagraphs dedicated to exo-planet detection requires careful and accurate optical modelling. Numerical representations of the propagation of aberrations through the system as well as simulations of the broadband wavefront compensation system using multiple DMs are critical when one devises an error budget for such a class of instruments. In this communication we introduce an analytical tool that serves this purpose for Phase Induced Amplitude Apodisation (PIAA) coronagraphs. We first start by deriving the analytical form of the propagation of a harmonic ripple through a PIAA unit. Using this result we derive the chromaticity of the field at any plane in the optical train of a telescope equipped with such a coronagraph. Finally we study the chromatic response of a sequential DM wavefront actuator correcting such a corrugated field and thus quantify the requirements on the manufacturing of PIAA mirrors.\\

\end{abstract}

\maketitle 

\section{Introduction}
%
%
Analytical propagation of wavefront errors through coronagraphs is the basis of all the recent error budget estimates for both ground and space based exo-planets imaging instruments. The literature has been particularly active in this area and our understanding of the sensitivity to aberrations of several coronagraphic solutions has considerably expanded over the past few years. For band limited coronagraphs, \citet{2005ApJ...634.1416S} first introduced an analytical propagator for low-order modes based on a Zernike decomposition. In parallel, \citet{2005ApJ...628..474S} carried out the same analysis for mid-spatial frequencies, based on a Fourier decomposition, comparing the sensitivities of band-limited coronagraphs. \citet{2003ApJ...596..702P} studied the impact of second order terms on the point spread function, a result revisited and formalized by \citet{AmirJosa}, who introduced the concept of frequency folding. \citet{AnandSensing} then focused on the propagation of mid-spatial frequencies through Apodised Pupil Lyot Coronagraphs. The case of out of pupil optics and the resultant chromatic phase to amplitude mixing was first tackled by  \citet{Stuart2DMsSPIE} for band-limited and shaped pupil coronagraphs and we recently provided a formal analytical approach that addresses this issue in \citet{2007ApJ...666..609P}.

Computing the sensitivity to aberrations proves to be more challenging task in the case of Phase Induced Amplitude Apodized (PIAA) coronagraph and for the Optical Vectorial Vortex Coronagraph (OVVC). PIAA coronagraphic technology, first introduced by Guyon \cite{2003A&A...404..379G}, is a promising solution since it makes most of the photons collected by the primary mirror of the telescope available for planet detection and characterization. This technique is based on two aspherical mirrors that redistribute the light in the pupil plane of a telescope so that it follows a given amplitude profile that leads to a Point Spread Function (PSF) having contrast levels close to $10^{-10}$. Because all the light collected is remapped using these mirrors, such a coronagraph has virtually no throughput loss. As a consequence, the angular resolution is undiminished and is close to the diffraction limit of $1 \lambda/D$, a feature comparable to the performances of a phase mask coronagraph such at the OVVC \cite{2009OExpr..17.1902M}, without any transmissive optics. The OVVC fully extinguishes on-axis starlight by introducing an azimuthal phase ramp at the focus of the coronagraph. This phase profile is obtained by manipulating the transverse polarization state of the light with space-variant birefringent elements \cite{2009OExpr..17.1902M}. The difficulty in modeling  OVVC resides in a proper treatment of the polarization effects and devising high fidelity models for manufacturing defects at the singularity located at the center of the focal plane mask. These issues are discussed and addressed in Mawet et.~al \cite{mawet:773914}.

The difficulty in PIAA modeling PIAA coronagraphs resides in devising accurate analytical models for the propagation of low-order and mid-spatial frequency aberrations. Since the optical surfaces of such a system are highly aspherical, classical tools based on the Fresnel approximation cannot be used. The first numerical diffractive study of wavefront propagation through such a coronagraph was carried out by \citet{2006ApJ...652..833B}, based on an expansion in Zernike polynomials. They showed that the high sensitivity to off-axis sources of pupil mappers was the cause of a higher sensitivity to low order aberrations. Herein we expand upon the results of \citet{Shaklan07SPIE} and derive a full treatment of the propagation of mid-spatial frequency harmonic aberrations through a two mirror pupil remapping system. Our main result is an analytical propagator for harmonic wavefront errors entering a PIAA coronagraph. The derivation of this analytical tool is presented in Section \ref{sec:propagator}.  The remainder of the paper illustrates how this propagator can be used to address the problem of broadband wavefront control for pupil mapping coronagraphs. Section \ref{sec:harmonic} follows the presentation of \citet{2007ApJ...666..609P} to derive an analytical expansion of the wavefront of a telescope equipped with such a coronagraph. Finally, in Section \ref{sec:broadband} we apply this analytical method to  predict the broadband performances of wavefront sensing and control systems applied to a pupil mapping coronagraph.

 %
 \section{Background}
As shown in Fig.~\ref{FigSetupPIAA}, a pupil remapping unit is composed of two aspherical mirrors that remap the light of an incoming pupil according to a prescribed apodisation profile \citet{2003ApJ...599..695T}. As presented in \citet{2003A&A...404..379G} a PIAA coronagraph is composed of two remapping units separated by a focal plane mask. The first set of aspherical mirror remaps the telescope pupil so that the starlight is concentrated in the core of a very high contrast Point Spread Function. The focal plane stop blocks the core of this PSF and hence removes the bulk of the starlight while preserving most of the photons form a potential companion. The purpose of the second remapping unit is to invert the remapping in order to restore the imaging properties of the whole apparatus. The purpose of this paper is to investigate the physics and propagation properties of a single two-mirrors remapping unit and thus solely focus on the ``forward'' combination shown on Fig.~\ref{FigSetupPIAA}. In a future communication we plan to use the findings reported here to study the performances of a full four mirror PIAA coronagraph. In this section we review the design equations of a circularly symmetric pupil remapper for which the designed apodization is independent of azimuth,
\begin{equation}
A(\tilde{x},\tilde{y}) = A(\tilde{r}).
\end{equation}
Here $(\tilde{x},\tilde{y})$ and $(\tilde{r},\tilde{\theta})$ are the location of the rays at $M2$ in Cartesian and polar coordinates respectively. Similarly, $(x,y)$ and $(r,\theta)$ are the location of rays at $M1$ in Cartesian and polar coordinates. In the most general case, as shown in the bottom right panel of Fig.~\ref{FigSetupPIAA}, the diffraction limited field at a location at M2, $(\tilde{x},\tilde{y})$, is given by the sum of diffracted wavelets at each point at the surface of M1.  In the particular case of ray optics there is no summation involved and there is only a one to one mapping between the field at $(\tilde{x},\tilde{y})$ and the field at the incident point on M1 $(x_0(\tilde{x},\tilde{y}),y_0(\tilde{x},\tilde{y}))$.  For a given $(\tilde{x},\tilde{y})$ the coordinates $(x_0(\tilde{x},\tilde{y}),y_0(\tilde{x},\tilde{y}))$ are derived from the mirror surfaces using Fermat's principle. $(r_0(\tilde{x},\tilde{y}),\theta_0(\tilde{x},\tilde{y}))$ are the coordinates of the same point in a polar system. Without loss of generality we  choose, for the remainder of this paper, to focus on pupil to pupil on-axis PIAA systems that are shown in the bottom two panels of  Fig.~\ref{FigSetupPIAA}.  Moreover if we choose $\theta_0 = \tilde{\theta}$, then, following the presentation of \citet{2003ApJ...599..695T}, the relationship between the location of the incident and outgoing rays can be written,
\begin{equation}
\frac{\partial r_0}{\partial  \tilde{r}} = \frac{\tilde{r}}{r_0} A(\tilde{r})^2.
\label{EqRemapPolar}
\end{equation}
where $r_0$ is the radial location of the incident ray at M1.  The design of the mirror shapes is then driven by the following coupled partial differential equations,
\begin{eqnarray}
\frac{\partial h}{\partial r}|_{r_0}& =& \frac{\tilde{r}(r_0)-r_0}{Z} \label{EqHR}\\
\frac{\partial \tilde{h}}{\partial \tilde{r}}|_{\tilde{r}}& =& \frac{\tilde{r}-r_0(\tilde{r})}{Z} \label{EqTildeHTildeR}
\end{eqnarray} 
where $h(r)$ is the height of M1, $h(\tilde{r})$ is the height of M2, and $Z$ is the distance between the two mirrors. Eq.~\ref{EqHR} and Eq.~\ref{EqTildeHTildeR} were derived using a ray optics approximation.  Unfortunately, in order to predict the contrast of an actual PIAA, it is necessary to carry out a full diffraction analysis.  

In particular, \citet{2006ApJ...636..528V} showed that edge propagation effects limit the contrast to at most $10^{-5}$ for a two mirrors PIAA coronagraph alone. \citet{2005astro.ph.12421P} subsequently showed that pre and post-apodisers can mitigate these effects and a $10^{-10}$ contrast can be recovered with little loss in throughput and angular resolution. Recently, we developed a new, purely analytical  approach to the diffraction problem \cite{PueyoThesis}. This new method can be applied to quantify the impact of diffraction of the edge of the remapping optics on contrast; using it we reproduced the results of \citet{2006ApJ...636..528V} which established a set of pre and post-apodisers that allow $10^{-10}$ with an aberration free PIAA \citet{pueyo:74400E}. Here we do not delve into such an analysis, which will be the object of a future detailed communication, and choose to focus on the analytical modeling of wavefront propagation. We start from the main theoretical result of \cite{PueyoThesis}. Namely, for any geometry, the diffraction limited field at $M2$ is well approximated by the following quadratic integral,
\begin{equation}
E_{out}(\tilde{x},\tilde{y}) = \frac{1}{i \lambda Z}\int_{M1} E_{in}(x,y) e^{i \frac{\pi}{\lambda Z} [(x-x_0)^2 + 2 (x-x_0) (y-y_0) + (y-y_0)^2]} dx \; dy.
\label{EqExpansionSecond}
\end{equation}
Where the ray optics remapping in cartesian coordinates is given by:
\begin{eqnarray}
x_0(\tilde{x},\tilde{y}) &=& r_0(\tilde{r}) \cos(\tilde{\theta})\\
y_0(\tilde{x},\tilde{y}) &=& r_0(\tilde{r}) \sin(\tilde{\theta})
\end{eqnarray}
The integral in Eq.~\ref{EqExpansionSecond} is a sum over all the contributions of each point at the surface of M1, while ray optics only relies on the field at $(x_0,y_0)$. Moreover, we show in App.~\ref{Sec:AppendixA} that for circularly symmetric systems the outgoing field is given by 
\begin{equation}
E_{out}(\tilde{r},\tilde{\theta}) = \frac{1}{i \lambda Z} \int_{M1} E_{in}(r,\theta)e^{i \frac{\pi}{\lambda Z} \left [ \frac{r_0(\tilde{r})}{\tilde{r} A(\tilde{r})^2} (r \cos(\theta -\tilde{theta}) - r_0(\tilde{r}) )^2 + \frac{\tilde{r}}{r_0(\tilde{r})} (r \sin(\theta -\tilde{theta}))^2 \right] } r dr \; d \theta.
\label{EqEllipseNonCentNonStreched}
\end{equation}
This result is the starting point for our analysis of the propagation of aberrations through PIAA systems. 

\section{Diffraction analysis of PIAA}\label{sec:propagator}

In this section we are interested in finding an analytical expression for the outgoing field when $E_{in}(x,y)$ is composed of harmonic ripples. While here we are considering the propagation of errors through a PIAA unit that is circularly symmetric we choose to describe the aberration in cartesian coordinates. By choosing a basis set that is not orthogonal over a circle we do not guarantee the unicity or the finite support of any given wavefront expansion. However this choice leads us to analytical insights about the wavefront mixing that occurs when a wavefront is propagated through a PIAA unit. Thus, here we follow the presentation of  \cite{Stuart2DMsSPIE}, and we study the propagation of harmonic ripples of the following form,
\begin{equation}
 e^{i \frac{2 \pi}{D} (N r \; \cos( \phi - \theta))} 
\end{equation}
where $N$ is the spatial frequency of the ripple and $\phi$ is its orientation. The propagation of such a complex disturbance through a circularly symmetric pupil mapping coronagraph is described by the following theorem.
\begin{theorem}
\label{ThmPropAbber}
\label{MainPropAbberThorem}
Consider a pupil mapping system with prescribed apodisation $A(\tilde{r})$, whose incident illumination is harmonic:
\begin{equation}
E_{in}(r,\theta) = e^{i \frac{2 \pi}{D} (N r \; \cos( \phi - \theta))} 
\end{equation}
Then, assuming that the edge effects are mitigated according to the methodology presented in \citet{2005astro.ph.12421P}, using pre and post apodisers, the field distribution at $M2$ is:
\begin{equation}
E_{out, N,\phi}(\tilde{r},\tilde{\theta}) = A(\tilde{r}) e^{i \frac{2 \pi}{D}N r_0(\tilde{r}) \cos (\tilde{\theta} -\phi)} e^{- i \frac{\pi \lambda Z N^2}{D^2} [\frac{\tilde{r} A(\tilde{r})^2}{r_0(\tilde{r})} cos^2 (\tilde{\theta}-\phi) + (\frac{r_0(\tilde{r})}{\tilde{r}})^2 sin^2 (\tilde{\theta}-\phi) ] }
\label{EqThPropAbber}
\end{equation}
Where $A(\tilde{r})$ is the apodisation profile that is induced by the two remapping mirrors.
\end{theorem}
Equation~\ref{EqThPropAbber} combines the geometrical and diffractive propagation effects of pupil mapping systems for mid-spatial frequencies aberrations. This theorem is derived using Eq.~\ref{EqEllipseNonCentNonStreched} with $E_{in} = e^{i \frac{2 \pi}{D} (mx+ny)}$, our purpose being to decompose any incoming wavefront before the remapping mirrors into a sum of Fourier harmonics and propagate them analytically.
%
\begin{proof}
We start with Eq.~\ref{EqEllipseNonCentNonStreched} and rewrite it into a form similar to the Fresnel integral of harmonic aberrations. First we express the harmonic ripple in polar coordinates,
\begin{eqnarray}
E_{out}(\tilde{r},\tilde{\theta})& =& \frac{1}{i \lambda Z} \int_{0}^{R} \int_{0}^{2 \pi} e^{i \frac{2 \pi}{D} N( r \cos \theta  \; cos \phi  +  r \sin \theta \; sin \phi)} e^{i \frac{\pi}{\lambda Z} \left [ \frac{r_0}{\tilde{r} A(\tilde{r})^2} (r \cos (\theta-\tilde{\theta}) - r_0 )^2 + \frac{\tilde{r}}{r_0} (r \sin (\theta-\tilde{\theta}))^2 \right] } r dr \; d \theta \nonumber \\
 &=& \frac{1}{i \lambda Z} \int_{M1} e^{i \frac{2 \pi}{D} N [u \cos (\tilde{\theta}- \phi)+v \sin (\tilde{\theta}- \phi)]} e^{i \frac{\pi}{\lambda Z} \left [ \frac{r_0}{\tilde{r} A(\tilde{r})^2} (u- r_0 )^2 + \frac{\tilde{r}}{r_0} v^2 \right] } du \; dv
\end{eqnarray}
where we have applied a series of coordinate rotations in order to write $E_{out}(\tilde{r},\tilde{\theta})$ in a form convenient to complete the square in the integrand. We then extract all the terms that do not depend on $(u,v)$ out of the integral and complete the squares. This yields
\begin{eqnarray}
E_{out}(\tilde{r},\tilde{\theta}) &=& \frac{1}{i \lambda Z}e^{i \frac{2 \pi}{D} N r_0 \cos (\tilde{\theta} - \phi)} e^{- i \frac{\pi \lambda z N^2}{D^2} [\cos^2(\tilde{\theta}-\phi)\frac{\tilde{r} A(\tilde{r})^2}{r_0} + sin^2(\tilde{\theta} - \phi) \frac{r_0}{\tilde{r}}]} \\
&& \times \int_{M1}e^{i \frac{\pi}{\lambda Z} \left [ \frac{r_0}{\tilde{r} A(\tilde{r})^2} (u- r_0 - \frac{\lambda z N}{D} \frac{\tilde{r} A(\tilde{r})^2}{r_0} \cos(\tilde{\theta}-\phi) )^2 + \frac{\tilde{r}}{r_0} (v-\frac{\lambda z N}{D} \frac{r_0}{\tilde{r}} \sin(\tilde{\theta}-\phi))^2 \right] } du \; dv. \nonumber
\end{eqnarray}
The first exponential factor corresponds to the ray optics remapping of the ripple. The second exponential factor accounts for propagation induced phase to amplitude conversion and the integral captures the edge oscillation effects due to the propagation, in the sense presented by \citet{2005astro.ph.12421P}. The propagation integral can be rewritten,
\begin{equation}
 \frac{1}{i \lambda Z} \int_{M1}e^{i \frac{\pi}{\lambda Z} \left [ \frac{r_0}{\tilde{r} A(\tilde{r})^2} (u- r_0 - \frac{\lambda z N}{D} \frac{\tilde{r} A(\tilde{r})^2}{r_0} \cos(\tilde{\theta}-\phi) )^2 + \frac{\tilde{r}}{r_0} (v-\frac{\lambda z N}{D} \frac{r_0}{\tilde{r}} \sin(\tilde{\theta}-\phi))^2 \right] } du \; dv = A(\tilde{r}) \int_{\mathcal{E}_{\lambda,\tilde{r},\tilde{\theta}}} e^{ i \rho^2} \rho d \rho d \psi
\end{equation}
where $\mathcal{E}_{\lambda,\tilde{r},\tilde{\theta}}$ is an ellipse centered at $( \frac{\lambda z N}{D} \frac{\tilde{r} A(\tilde{r})^2}{r_0} \cos(\tilde{\theta}-\phi) )^2, \frac{\lambda z N}{D} \frac{r_0}{\tilde{r}} \sin(\tilde{\theta}-\phi))^2)$ of semi major axis $ a_{\mathcal{E}} = (\frac{\pi}{\lambda Z} \frac{r_0}{\tilde{r} A(\tilde{r})^2} ) ^{-1}$ and semi minor axis $ b_{\mathcal{E}} = (\frac{\pi}{\lambda Z} \frac{\tilde{r}}{r_0} ) ^{-1}$. When the phase oscillations at $M2$ have been mitigated using pre and post-apodisers, along the lines of \citet{2005astro.ph.12421P},  the diffractive properties are equivalent to a ray optics propagation, namely $ a_{\mathcal{E}} \rightarrow \infty $ and $ b_{\mathcal{E}} \rightarrow \infty $
\begin{equation}
 \frac{1}{i \lambda Z} \int_{M1}e^{i \frac{\pi}{\lambda Z} \left [ \frac{r_0}{\tilde{r} A(\tilde{r})^2} (u- r_0 )^2 + \frac{\tilde{r}}{r_0} v^2 \right] } du \; dv \simeq A(\tilde{r}).
\end{equation}
This finishes our proof.
\end{proof}
In the remainder of the paper we use Theorem 1 to compute the effects of propogated harmonic aberrations on the final image and to determine limits on the wavefront control system.
Another application of Eq.~\ref{EqThPropAbber} is presented in App.~\ref{Sec:AppendixB} where we evaluate the sensitivity of PIAA systems to off-axis sources.

\section{Propagation of harmonic aberrations}\label{sec:harmonic}
In a previous paper \cite{2007ApJ...666..609P} we illustrated the impact of out of pupil plane optic on the final aberrated field. These are responsible for mixing the amplitude/phase nature of aberrations and for changing their chromatic behavior. We showed that these effects could be mitigated using a two sequential DM wavefront controller. The same problem arises for PIAA coronagraphs, since it is intrinsically composed of two surfaces that are not only out of conjugacy with respect to the telescope pupil, but that also are highly non-parabolic, thus introducing extra perturbations on the wavefront. In this section we show how to use the propagator derived in Eq.~\ref{EqThPropAbber} to quantify the chromaticity of post-PIAA aberrated wavefronts. 
\subsection{Phase to amplitude conversion}
We first consider the case of a phase error in the field right after M1 that is given by $E_{in}(x,y)  = e^{i \frac{2\pi}{\lambda} h(h,y)}$. Note that this phase error can either stem from optics before PIAA M1 or on the surface of M1,
\begin{eqnarray}
h (x,y) &=& \sum_{m,n} \lambda_0 b_{m,n} e^{i \frac{2 \pi}{D} (m x+ ny)} \\
&=& \sum_{m,n} \lambda_0 b_{m,n} e^{i \frac{2 \pi}{D} \sqrt{m^2+n^2}( x \; \cos \theta_{n,m}+ y \; \sin \theta_{m,n} )} \label{dh_polar}
\end{eqnarray}
where $\theta_{m,n} = \tan^{-1}(\frac{n}{m})$, $\lambda_0$ is the central wavelength and $b_{-m,-n} = b^{\ast}_{m,n}$ are non dimensional Fourier coefficients. We assume for now that the  $b_{m,n}$ coefficients are small enough that the field at $M1$ can be approximated by
\begin{equation}
e^{i \frac{2 \pi}{\lambda} h } \simeq  (1+i \frac{2 \pi}{\lambda} h ). 
\end{equation}
%
%
%
Note that under this linear approximation, a phase errors is equivalent to an imaginary disturbance of the pupil plane and an amplitude error to a real disturbance. Assuming a circularly symmetric PIAA,  Eq.~\ref{EqEllipseNonCentNonStreched} yields,
\begin{equation}
E_{out}(\tilde{r},\tilde{\theta})= \frac{1}{i \lambda Z} \int_{M1} (1+ i \frac{2 \pi}{\lambda} h(x,y)) e^{i \frac{\pi}{\lambda Z} \left [ \frac{r_0}{\tilde{r} A(\tilde{r})^2} (r \cos (\theta-\tilde{\theta}) - r_0 )^2 + \frac{\tilde{r}}{r_0} (r \sin (\theta-\tilde{\theta}))^2 \right] } r dr \; d \theta. \\
\end{equation}
Therefore, using the Fourier expansion of $\delta h(x,y)$ in Eq.~\ref{dh_polar} and Thm.~\ref{ThmPropAbber},
\begin{eqnarray}
E_{out}(\tilde{r},\tilde{\theta}) = A(\tilde{r}) \big( 1+  \frac{2 \pi \lambda_0}{\lambda}  \sum_{m,n} b_{m,n} e^{i \frac{2 \pi}{D} \sqrt{m^2+n^2} (r_0 \cos \tilde{\theta} \; \cos \theta_{m,n} + r_0 \sin \tilde{\theta} \; \sin \theta_{m,n} )} \nonumber \\
e^{-i \frac{\pi \lambda Z (m^2+n^2)}{D^2} [(\frac{\tilde{r} A(\tilde{r})}{r_0})^2 \cos^2 (\tilde{\theta}-\theta_{m,n}) + (\frac{r_0}{\tilde{r}})^2 \sin^2 (\tilde{\theta}-\theta_{m,n})]} \big).
\end{eqnarray}
The general form for the propagator of phase errors in classical coronagraph behaves, as derived in \citet{2007ApJ...666..609P}, as $e^{- i \frac{\pi \lambda z N^2}{D^2}}$. The propagator derived here does feature the same behavior but also captures the high curvature of the PIAA optics through the angular magnification factor $(\frac{\tilde{r} A(\tilde{r})}{r_0})^2 \cos^2 (\tilde{\theta}-\theta_{m,n}) + (\frac{r_0}{\tilde{r}})^2 \sin^2 (\tilde{\theta}-\theta_{m,n})$. This magnification is different in the radial and tangential directions. Therefore, in the angular spectrum factor, the contribution of the radial magnification is weighted by the relative orientation of the ripple, $\theta_{m,n}$, with respect to the line of observation, $\tilde{\theta}$. This yields a term in $\cos^2(\tilde{\theta}-\theta_{m,n})$.  The same consideration for the tangential direction yields a term in $\sin^2(\tilde{\theta}-\theta_{m,n})^2$. Seen from $M2$ it is as if the ripples at $M1$ were propagating along two orthogonal pupil mappers of different linear magnification laws.
As a consequence, a phase ripple at $M1$ will not only appear at $M2$ as a condensed oscillatory pattern, as predicted by the laws of geometric optics, but will also see some of its energy transfered to amplitude, thus creating phase induced amplitude errors. This effect gets stronger with spatial frequency as illustrated in Figs.~\ref{figPhaseInducedAmplitudeN6} and \ref{figPhaseInducedAmplitudeN10}. In Fig.~\ref{figPhaseInducedAmplitudeN10}, computed at $\lambda = 700$ nm, for a spatial frequency of $10$ cycles per aperture, a pupil diameter of $D = 3$ cm and a mirror separation of $z = 1$ m, the effective propagation distance is equivalent to a quarter of a Talbot distance and the conversion is total: all the wavefront error becomes amplitude. This phase to amplitude conversion behaves as $D^2/ \lambda z$, the Fresnel number of the PIAA unit. Thus this design parameter has a direct impact on the feasibility of broadband wavefront control.  \\
\subsection{Amplitude induced phase error}
The same approach can be carried out starting with amplitude errors in the field right after $M1$,
\begin{equation}
E_{in}(x,y) =r_0 (1 + \sum_{m,n} a_{m,n} e^{i \frac{2 \pi}{D} (m x+ ny)} )
\end{equation}
with $a_{-m,-n} = a_{m,n}^{\ast}$ are non dimensional coefficients and $r_0$ is the average transmissivity of the incident field. This leads to the derivation of amplitude induced phase errors that is given by 
\begin{eqnarray}
E_{out}(\tilde{r},\tilde{\theta}) = A(\tilde{r}) \big( 1+ \sum_{m,n} a_{m,n} e^{i \frac{2 \pi}{D} \sqrt{m^2+n^2} (r_0 \cos \tilde{\theta} \; \cos \theta_{m,n} + r_0 \sin \tilde{\theta} \; \sin \theta_{m,n} )} \nonumber \\
e^{-i \frac{\pi \lambda Z (m^2+n^2)}{D^2} [(\frac{\tilde{r} A(\tilde{r})}{r_0})^2 \cos^2 (\tilde{\theta}-\theta_{m,n}) + (\frac{r_0}{\tilde{r}})^2 \sin^2 (\tilde{\theta}-\theta_{m,n})]} \big).
\end{eqnarray}
This amplitude to phase conversion is illustrated in Figs.~\ref{figAmplitudeInducedPhaseN6} and \ref{figAmplitudeInducedPhaseN10},computed at $\lambda = 700$ nm. Once again, for a spatial frequency of $10$ cycles per aperture, a pupil diameter of $D = 3$ cm and a mirror separation of $z = 1$ m, we observe that for a point located at the center of $M2$, the effective propagation distance is equivalent to a quarter of a Talbot distance and the conversion is total: all the wavefront error becomes phase. These considerations raise a fundamental issue when one seeks to create a broadband null in the image plane of a PIAA coronagraph. Because this phase to amplitude mixing is chromatic, it alters the bandwidth of wavefront correctors. This is the question we address next by deriving a full expansion of the chromaticity of the wavefront after a PIAA coronagraph.

\subsection{Wavelength expansion of the propagated electrical field}
In the previous subsections we derived the field propagation for a harmonic aberration  at a single wavelength. Here we seek an expansion for the propagated field over a band of wavelengths.  Our main result is stated in the following theorem.
\begin{theorem}
\label{theoexpansionPIAA}
We assume that the edge effects are mitigated in the PIAA unit via pre and post-apodisers and that the optics before PIAA $M1$ and after PIAA $M2$ are such that  $m \; n \ll \frac{D}{\sqrt{ \lambda z}}$, where $D$ is the optic diameter and $z$ is the distance between a given optics and the conjugate of the telescope pupil. Then, the field in any plane of the optical train of a telescope equipped with a PIAA coronagraph, after an arbitrary number of reflections on aberrated optics, and a propagation through the two mirrors remapping unit, can be expanded using the following $\lambda$-Fourier expansion
\begin{equation}
E(\tilde{x},\tilde{y}) = A(\tilde{x},\tilde{y}) \left(1+\sum_{m,m} \sum_k i^k \frac{f_{m,n}^{-k} \lambda_0^k}{\lambda^k} e^{i \frac{2 \pi}{D} (m \tilde{x}+n \tilde{y})}\right)
\label{eqtheorealPIAA}
\end{equation}
where $f_{-m,-n}^{-k} = (f_{m,n}^{-k})^{*}$. That is, the odd terms in the wavelength expansion are imaginary and the even terms are real.
\end{theorem}
Note that this theorem is exactly the same as Theorem 2 in \citet{2007ApJ...666..609P}. It is as if for wavefront mixing purposes, a PIAA coronagraph behaves analytically exactly like a classical coronagraph; except that the chromatic phase/amplitude conversion due to propagation is magnified by a factor of $\mathcal{M}$, the angular magnification of the PIAA unit. This effect will be quantified in the next section.
\begin{proof}
In \citet{2007ApJ...666..609P} we established, using an induction argument and assuming that the optical surfaces are all parabolic or flat, that Eq.~\ref{eqtheorealPIAA} was true in any plane of a classical coronagraph as long as $m , n \ll \frac{D}{\sqrt{ \lambda z}}$ . Here we are interrested in proving that the propagation through the non-parabolic optics of a PIAA coronagraph conserves this property: namely if the field $E_{in}$ at $M1$ is such that Eq.~\ref{eqtheorealPIAA} is true, then the field $E_{out}$ after $M2$ also satisfies this property. We write the field at $M1$ as:
\begin{equation}
E_{in}(x,y) = 1+\sum_{m,m} \sum_k i^k \frac{f_{m,n}^{M1,-k} \lambda_0^k}{\lambda^k} e^{i \frac{2 \pi}{D} (m x+n y)}
\end{equation}
As a result of theorem \ref{MainPropAbberThorem}, the field at $M2$ is:
\begin{eqnarray}
E_{out}(\tilde{r},\tilde{\theta}) &=&A(\tilde{r}) \big[ 1+\sum_{m,n} \sum_k i^k \frac{f_{m,n}^{M1,-k} \lambda_0^k}{\lambda^k}  e^{i \frac{2 \pi}{D}N_{m,n} (r_0 \cos \tilde{\theta} \; \cos \phi_{m,n} + r_0 \sin \tilde{\theta} \; \sin \phi_{m,n} )} \nonumber \\
& &  e^{-i \frac{\pi \lambda Z N_{m,n}^2}{D^2} [(\frac{\tilde{r} A(\tilde{r})^2}{r_0})^2 \cos^2 (\tilde{\theta}-\phi_{m,n}) + (\frac{r_0}{\tilde{r}})^2 \sin^2 (\tilde{\theta}-\phi_{m,n}) ] } \big]
\label{Eq:PropAtM2Basic}
\end{eqnarray}
We write the propagator at a spatial frequency $(m,n)$ as:
\begin{equation}
e^{-i \frac{\pi \lambda Z N_{m,n}^2}{D^2} [(\frac{\tilde{r} A(\tilde{r})^2}{r_0})^2 \cos^2 (\tilde{\theta}-\phi_{m,n}) + (\frac{r_0}{\tilde{r}})^2 \sin^2 (\tilde{\theta}-\phi_{m,n}) ] } = e^{- i \frac{\lambda}{\lambda0} \psi_{m,n}(\tilde{r},\tilde{\theta})}
\end{equation}
If we expand this exponential in a Taylor series then Eq.~\ref{Eq:PropAtM2Basic} becomes:

\begin{eqnarray}
E_{out}(\tilde{r},\tilde{\theta}) &=&A(\tilde{r}) \big[ 1+\sum_{m,n} \sum_k \sum_{p = 0}^{\infty} i^{k-p} \frac{f_{m,n}^{M1,-k} \lambda_0^{k-p}}{ p ! \lambda^{k-p}}  e^{i \frac{2 \pi}{D}N_{m,n} (r_0 \cos \tilde{\theta} \; \cos \phi_{m,n} + r_0 \sin \tilde{\theta} \; \sin \phi_{m,n} )} \nonumber \psi_{m,n}(\tilde{r},\tilde{\theta})^p \big] \nonumber \\
E_{out}(\tilde{r},\tilde{\theta}) &=&A(\tilde{r}) \big[ 1+ \sum_k \sum_{p = 0}^{\infty} i^{k-p} \frac{\lambda_0^{k-p}}{\lambda^{k-p}} \sum_{m,n} \frac{f_{m,n}^{M1,-k}}{p!}  \psi_{m,n}(\tilde{r},\tilde{\theta})^p e^{i \frac{2 \pi}{D}N_{m,n} (r_0 \cos \tilde{\theta} \; \cos \phi_{m,n} + r_0 \sin \tilde{\theta} \; \sin \phi_{m,n} )}  \big] \nonumber \\
\end{eqnarray}
We then re-write as a new Fourier series the function of $(\tilde{r},\tilde{\theta})$, that is represented by the sum over $(m,n)$ on the right of the $\frac{\lambda_0^{k-p}}{\lambda^{k-p}}$ factor. Since we have the freedom to arbitrarily index these new Fourier coefficients, for clarity we choose to call them $f_{m',n'}^{M2,k,p} = f_{m',n'}^{M2,-(k - p)}$
\begin{eqnarray}
& &\sum_{m,n} \frac{f_{m,n}^{M1,-k}}{p!}  e^{i \frac{2 \pi}{D}N_{m,n} (r_0 \cos \tilde{\theta} \; \cos \phi_{m,n} + r_0 \sin \tilde{\theta} \; \sin \phi_{m,n} )} \psi_{m,m}(\tilde{r},\tilde{\theta})^p \nonumber \\
& =&  \sum_{m',n'} f_{m',n'}^{M2,-(k - p)}e^{i \frac{2 \pi}{D}N_{m',n'} (\tilde{r} \cos \tilde{\theta} \; \cos \phi_{m',n'} + \tilde{r} \sin \tilde{\theta} \; \sin \phi_{m',n'} )}. 
\end{eqnarray}
The $ f_{m',n'}^{M2,-(k - p)}$ coefficients at $M2$ can be written as
\begin{equation}
f_{m',n'}^{M2,-(k - p)} = \int \int \sum_{m,n} \frac{f_{m,n}^{M1,-k}}{p!}  \psi_{m,n}(\tilde{r},\tilde{\theta})^p e^{i \frac{2 \pi}{D}(m x_0(\tilde{x},\tilde{y}) + n y_0(\tilde{x},\tilde{y}) )} e^{- i \frac{2 \pi}{D} (m' \tilde{x} + n' \tilde{y})} d \tilde{x} \; d \tilde{y}
\label{Eq:FourCoeffs}
\end{equation}
where we chose to represent the integrand in cartesian coordinates for clarity. Finally we reduce the double sum over $(k,p)$ to a single sum since their argument only depends on the difference $k' =k-p$. This yields
\begin{equation}
E_{out}(\tilde{x},\tilde{y}) = A(\tilde{x},\tilde{y}) \left(1+\sum_{m',n'} \sum_{k'} i^{k'} \frac{f_{m',n'}^{M2,-k'} \lambda_0^{k'}}{\lambda^{k'}} e^{i \frac{2 \pi}{D} (m' \tilde{x}+n' \tilde{y})} \right),
\label{EM2_expansion}
\end{equation}
which finishes the proof, since for optics after the PIAA unit we can use the results of \citet{2007ApJ...666..609P}.
\end{proof}

Note that Eq.~\ref{Eq:FourCoeffs} establishes an explicit relationship between the Fourier coefficients of electrical field distributions before and after a PIAA. Such a relationship is useful in wavefront control applications. As mentioned earlier, the only difference between classical coronagraphs and PIAA is that for a PIAA the errors are propagated through a modified angular spectrum.  The result is that the phase to amplitude mixing depends on the location on $M2$ and is stronger/weaker than for a classical coronagraph by a factor of $(\frac{\tilde{r} A(\tilde{r})}{r_0})^2 \cos^2 (\tilde{\theta}-\phi_{m,n}) + (\frac{r_0}{\tilde{r}})^2 \sin^2 (\tilde{\theta}-\phi_{m,n})$. This location dependent wavefront mixing impacts the broadband performance of DM based wavefront controllers.  It drives the size of the dark zone achievable using such controllers, which we discuss next.
\subsection{Largest correctable spatial frequency}
%
%
When the  spatial frequency is small enough, the largest two terms in the expansion in Eq.~\ref{EM2_expansion} are $i f_{m,m}^{M2,-1}/\lambda$, phase errors, and $f_{m,n}^{M2,0}$, amplitude errors. However, when the spatial frequency of the aberration gets larger, the propagated terms become larger and higher order wavelength dependent terms grow. If two sequential DMs that follow the PIAA unit are used to correct the phase errors, $i \; f_{m,m}^{M2,-1}/\lambda$ term, and the amplitude errors, $f_{m,n}^{M2,0}$, a residual halo appears at high-spatial frequencies due to these higher order terms. Because of the $N^2$ dependence of the angular spectrum propagator this halo is a strongly increasing function of spatial frequency. In \citet{2007ApJ...666..609P} we defined the highest correctable spatial frequency, $N_{limitC},$ as the spatial frequency of a ripple for which the application of a two sequential DM correction no longer provides better broadband contrast than for the case without wavefront correction. Ripples above $N_{limitC}$ cannot be corrected over a broadband because the chromatic mixing of the wavefront is too large. For a classical coronagraph, designed using parabolic optics, this limiting spatial frequency can be written as follows,
\begin{equation}
N_{limitC} = \frac{D} {\sqrt{\lambda_0 z}}(\frac{\lambda_0}{\Delta \lambda})^{1/2}.
\label{EqLimitSFClassic}
\end{equation}
The same considerations are valid for a PIAA coronagraph. However, when the DMs are located after the remapping mirrors, then the average spatial frequency seen by the propagator is increased by a factor of $\mathcal{M}$, as shown in Eq.~\ref{Eq:FourCoeffs} where the Fourier kernel is now written as a function of $(x_0(\tilde{x},\tilde{y}),y_0(\tilde{x},\tilde{y}) )$. Thus, if the maximal correctable spatial frequency is expressed in terms of cycles per aperture before the PIAA, then $N_{limitPIAA}$ becomes
\begin{equation}
N_{limitPIAA} = \frac{D}{\mathcal{M} \sqrt{\lambda_0 z}}(\frac{\lambda_0}{\Delta \lambda})^{1/2}
\label{EqLimitSFPIAA}
\end{equation}
While the final wavefront expansion is similar for PIAA and classical coronagraphs, the higher order propagation terms are larger for PIAA, thus reducing the largest spatial frequency correctable under a broadband illumination. With DMs located before the PIAA unit the largest correctable spatial frequency would be driven by the surface errors at $M2$, back-propagated to the plane of $M1$. We will study this configuration in a future communication. Here we emphasize that a two sequential DM wavefront controller that follows a PIAA unit will not be able to correct spatial frequencies above $N_{limitPIAA}$, expressed in Eq.~\ref{EqLimitSFPIAA}, over a broadband, due to a phase to amplitude mixing that is too strong.


In the case studied here, a PIAA that does not include de-mapping mirrors after the focal plane stop \cite{2003A&A...404..379G}, the image plane contribution of aberrations above $N_{limitPIAA}$ extends all the way to low spatial frequencies. This low spatial frequency leakage can be explained by using Fig.~\ref{figPhaseInducedAmplitudeN6} to Fig.~\ref{figAmplitudeInducedPhaseN10}: at the edges of $M2$ the aberration appears mostly as a small spatial frequency ripple and consequently throws light near the core of the PSF.  While considerably damped by the apodisation profile at $M2$, such a leakage corresponds to a very chromatic wavefront that is highly uncorrectable under a broadband illumination using two DMs in series after the PIAA unit. In the case of classical coronagraphs, aberrations above $N_{limit}C$ only have a small impact, due to the tail of the airy function, on the contrast in the Dark Hole of the coronagraph. In the case of a PIAA that does not include de-mapping mirrors after the focal plane stop, because of this low spatial frequency leak, aberrations above $N_{limitPIAA}$ can potentially have an impact at low spatial frequencies and thus influence the performances of the wavefront control system. In the next section we present, as an illustration of the analytical propagator derived above, numerical simulations that quantify the effect of the chromatic wavefront mixing on the overall post-correction contrast in the following configuration: no de-mapping mirrors and a two sequential DM wavefront actuator that follows the PIAA unit. 


\section{Contrast predictions}
\label{sec:broadband}
 We use the analytical expansion in Eq.~\ref{eqtheorealPIAA} to predict the best broadband contrast that can be achieved by a PIAA unit in the presence of fixed wavefront errors compensated by a pair of sequential DMs located after the remapping mirrors. Note that different, and potentially better, broadband performances can be obtained with DMs located before the remapping mirrors or a de-mapping unit after the focal plane mask. While these architectures can be studied using the approach presented here, their implementation requires novel wavefront control algorithms that are beyond the scope of this paper. Thus we decided here to focus on quantifying the limitations of the simplest solution possible. We will extend this study to all possible combinations of DM before / after the PIAA mirrors and with /without de-mapper in a future communication.
 
  Because the propagator is wavelength dependent, the wavelength expansion of the field at $M2$ exhibits an infinite number of term, whereas, as shown by \cite{2006ApOpt..45.5143S}, two sequential DMs can only correct for the $\lambda^0$ and $i \;1/\lambda$ terms. In this section we isolate one fourier component, at a given spatial frequency, and quantify how well a two sequential DMs wavefront controller can reject it under a broadband illumination. 
 \subsection{Methodology}
For a given phase error at $M1$, and a bandwidth centered around $\lambda_0$, we use a first order expansion of the wavefront
\begin{equation}
 E_{in} = 1 + i \frac{\lambda_0}{\lambda} e^{i \frac{2 \pi}{D} (m x + ny) }.
\label{LinAbber}
\end{equation}
We propagate the ripple according to Eq.~\ref{EqThPropAbber}.  We repeat this process for each wavelength in the band considered and thus build a data cube of fields at $M2$ that is represented on the left panel of Fig.~\ref{figPupilCubes}. Note that here with this first order expansion we are only studying the effect of PIAA propagation on small wavefront errors, leaving aside non-linear effects due to wavefront excursions. 
We then compute the electrical field distribution at the final image plane for each wavelength. We are interested in an annulus between the $IWA$ and the $OWA$ in  $(\lambda/D)_{On Sky}$, as represented in the right panel of  Fig.~\ref{figPupilCubes}, that is the region where we expect the wavefront control system to operate and create a Dark Hole. For the simulations shown here we chose $IWA =2 \; (\lambda/D)_{On Sky}$ and $OWA = 7.5 \; (\lambda/D)_{On Sky}$. These angles correspond to angular separation on the sky and are related to the actual units at the science focal plane by  $ (\lambda/D)_{Camera} = \mathcal{M}  (\lambda/D)_{On Sky}$, where $\mathcal{M}$ is the angular magnification of the PIAA unit.  Because the PSF of single spatial frequency ripples that have been propagated through a PIAA is extended, as shown on Fig.~\ref{figOffAxisPSFN24}, a ripple such that $\sqrt{m^2 +n^2} > OWA$ will still leak in the Dark Hole. The calculation we are carrying out in this section quantifies how much this leak is correctable using two DMs in series after the PIAA unit. To do so we proceed as follows. Assume that the field at a given pixel of the image plane is $E_{\lambda}^{(Image)}(\xi,\eta) = E_{\lambda}^{Re}(\xi,\eta) + i E_{\lambda}^{Im}(\xi,\eta)$. Then we assume that a perfect dual DM wavefront controller affects this field in the following fashion:
\begin{equation}
E_{\lambda}^{(Image - DMs)}(\xi,\eta) = (E_{\lambda}^{Re}(\xi,\eta) - E_{\lambda_0}^{Re}(\xi,\eta))  + i (E_{\lambda}^{Im}(\xi,\eta) - \frac{\lambda_0}{\lambda} E_{\lambda_0}^{Im}(\xi,\eta)).
\label{EqCancelDM}
\end{equation}
That is, we are assuming that the ideal wavefront controller cancels the electrical field at the central wavelength and features an achromatic leak for the real terms and a $1/\lambda$ leak for the imaginary terms. Another possibility would be to assume a controller that features the same chromatic dependance but that minimizes the chromatic residual intensity over the entire spectral bandwidth. The outcomes of both approaches are similar in terms of broadband performances and thus here we focus on the one described by Eq.~\ref{EqCancelDM}, which is illustrated in Fig.~\ref{LambdFitTwoOrders}. In this example the optical design is such that $D = 3$ cm, $z = 1$ m, $\lambda = 700$ nm and the spatial frequency $N = 7$ cycles per aperture. These values lead to a chromatic mixing of the wavefront that is so strong that even after an ideal wavefront controller, the contrast under a $10$ percent bandwidth is still $10^{-6}$ at its worse. Since we have assumed a ripple of amplitude $1$ we can conclude that, for the PIAA design used in the example of Fig.~\ref{LambdFitTwoOrders}, phase errors of spatial frequency $7$ cycles per aperture before $M1$ cannot create speckles that are larger than $10^{-4}$ in order to be corrected by a dual DMs wavefront controller located after the PIAA unit. Next we repeat this approach for a variety of spatial frequencies, optics size and separations and bandwidths. 
 \subsection{Results}
The first parameter studied here is the speckle extinction as a 
function of spatial frequency and Fresnel number. Because propagation 
effects scale with the Fresnel number,  $\mathcal{F}=\frac{D^2}{\lambda z}$, we expect the brodband residual halo due to 
higher order terms to increase when the Fresnel number decreases. Fig.~\ref{ContVSSfVSPupSize} illustrates this 
feature. It shows how the maximum of a composite PSF over several 
wavelengths, after the application of an ideal wavefront control, behaves as a function of spatial frequency and Fresnel number. For this figure we used a dark zone going from $IWA =2 \; (\lambda/D)_{On Sky}$ to $OWA = 7.5 \; (\lambda/D)_{On Sky}$. This OWA correspond to the $32 (\lambda/D)_{Camera}$ outer limit due to the 
 limited number of degree of freedom of a DM with $64$ actuators across the pupil, divided by the magnification, $\mathcal{M}$, of the pupil mapping unit considered here. Fresnel number is a crucial parameter when designing PIAA units: as seen on Fig.~\ref{ContVSSfVSPupSize}, for $ \mathcal{F} = 11250 $ a two sequential DMs wavefront controller manages to extinguish speckles by seven to eight orders of magnitude over a broad range of mid-spatial frequencies. Such a level of extinction coupled with reasonably small wavefront errors to start with, makes broadband wavefront control over a dark hole in the image plane with a PIAA coronagraph feasible. However for $\mathcal{F} = 140$, this extinction is reduced to two or three orders of magnitude, which considerably hampers prospects for broadband wavefront control with such a design.
The second parameter studied is the speckle 
extinction as a function of spatial frequency and bandwidth.
Fig.~\ref{ContVSSfVSBandwidth} plots the worst contrast in the same dark zone as a function of the spatial frequency of the input ripple with each curve corresponding to 
bandwidths of $\Delta \lambda / \lambda = 0.1, \; 0.2, \; 0.3$. In this example, for $\mathcal{F} = 11250$, the bandwidth does not have much of an effect on broadband contrast. However, as seen on Fig.~\ref{ContVSSfVSBandwidth32Legend}, with $\mathcal{F} = 1250$ the propagation effects of high spatial frequencies aberrations through the PIAA coronagraph become very  large, and, as expected, the performance of the 2 DM controller becomes sensitive to bandwidth.

This analysis provides a methodology for deriving prescriptions on the optics that precede a PIAA coronagraph, when the DMs are located after the pupil remapping unit. Our main conclusion is that for Fresnel number larger than $10^4$, chromatic propagation effects do not have much overall impact on the effectiveness of a two DM wavefront controller. Indeed, as shown in Fig.~\ref{ContVSSfVSBandwidth},  such an apparatus manages to cancel a given spatial frequency by $8$ orders of magnitude over several bandwidths ranging from ten to thirty percent. In this regime the broadband halo is more likely to be dominated by non-linear effects such as frequency folding, that scales as $1/ \lambda^2$, and is only partially correctable using two sequential DMs. This effect has been discussed in previous communications \citet{AmirJosa} and they have not been taken into account in the present paper. Nevertheless, since propagation effects become larger when $\mathcal{F}$ get smaller, this study shows that in the regime of $\mathcal{F} \simeq 100$, chromaticity of the residual halo is driven by the propagated wavefront errors from $M1$ to $M2$. For instance, for a PIAA designed with $\mathcal{F} = 140$, a $10^{-10}$ contrast over a $20$ percent bandwidth seems an unachievable goal with DMs located after the PIAA unit. Indeed, as shown on Fig.~\ref{ContVSSfVSBandwidth32Legend}, the pre-wavefront control speckles due to ripples of $10$ to $50$ cycles per aperture before $M1$ would need to be small enough so that the raw contrast is lower than $10^{-7}$ in order to be correctable to the $10^{-10}$ level with a post PIAA dual DM wavefront controller.
\section{Closing remarks and future work}
In this paper we derived an analytical propagator for aberrations through a PIAA coronagraph. This propagator, is based on a Fourier expansion, and captures the wavelength dependence of the field after the coronagraph. This aspect is of critical importance with respect to the design of ongoing and future experiments based on this coronagraph.  It provides a technique for tolerancing and error budgeting coronagraphic optics when PIAA is used in conjunction with a broadband wavefront controller. In the case of two DMs located after the two remapping mirrors, we applied a contrast evaluation procedure which predicts that broadband wavefront control for this architecture is only possible for PIAA Fresnel numbers that are larger than $2000$. In the near future we will explore algorithms that control two sequential DMs that are located before $M1$, and study the case of a PIAA coronagraph with de-mapping mirrors, such architectures potentially being more favorable and allowing smaller PIAA Fresnel numbers.

\appendix

\section{Second order expansion of the PIAA integral with circular symmetry}
\label{Sec:AppendixA}
We start from Eq.~\ref{EqExpansionSecond} and write the expression of the partial derivatives of the remapping, Eq.~\ref{EqRemapPolar}, in polar coordinates, 
\begin{eqnarray}
\frac{\partial \tilde{x}}{\partial x_0} &=& \frac{\partial \tilde{r}}{\partial r_0} cos^2 \theta_0+ \frac{\tilde{r}}{r_0} sin^2 \theta_0 \nonumber \\
\frac{\partial \tilde{y}}{\partial y_0} &=& \frac{\partial \tilde{r}}{\partial r_0} sin^2 \theta_0+ \frac{\tilde{r}}{r_0} cos^2 \theta_0   \label{EqChainBis} \\
\frac{\partial \tilde{x}}{\partial y_0} &=&  (\frac{\partial \tilde{r}}{\partial r_0}  - \frac{\tilde{r}}{r_0}) cos \theta_0 \; sin \theta_0 \nonumber
\end{eqnarray} 
where we have assumed that $\tilde{\theta} = \theta_0$. Eqs.~\ref{EqChainBis} corresponds to the partial derivatives of the inverse remapping and are derived using the chain rule. We are interested in changing the integration variables in Eq.~\ref{EqExpansionSecond} from cartesian to polar coordinates:
\begin{eqnarray}
(x-x_0)^2 &=& r^2 cos^2 \theta - 2 r r_0 cos \theta \; cos \theta_0 + r_0^2 cos^2 \theta_0 \nonumber \\
(y-y_0)^2 &=& r^2 sin^2 \theta - 2 r r_0 sin \theta \; sin \theta_0 + r_0^2 sin^2 \theta_0 \label{EqCartToPol}   \\
(x-x_0) (y - y_0) &=& r^2 cos \theta \; sin \theta -  r r_0 (sin \theta \; cos \theta_0+cos \theta \; sin \theta_0) + r_0^2 cos \theta_0 \; sin \theta_0 \nonumber \\
dx \; dy & = & r dr \; d \theta \nonumber
\end{eqnarray}
where the integral is taken over a circle of radius $R$ and centered at the center of this circle. We will call such a domain of integration $\mathcal{C}_{(0,0)}^R$. After some algebraic manipulations we can find a simple expression for the radial terms in the exponential factor of the quadratic expansion:
\begin{eqnarray}
r^2 &:& \frac{d \tilde{r}}{d r_0} \cos^2 (\theta - \theta_0) + \frac{\tilde{r}}{r_0} \sin^2 (\theta  -\theta_0) \nonumber\\
r_0^2 &:&  \frac{d \tilde{r}}{d r_0} \label{EqTermExpCirc} \\
r r_0 &:& 2  \frac{d \tilde{r}}{d r_0} \cos (\theta - \theta_0) \nonumber
\end{eqnarray}
As a consequence the radial field distribution after $M2$ becomes:
\begin{eqnarray}
E_{out}(\tilde{r}) &=& \frac{1}{i \lambda Z} \int_{\mathcal{C}_{(0,0)}^R} e^{i \frac{\pi}{\lambda Z} \left [ \frac{r_0}{\tilde{r} A(\tilde{r})^2} (r cos \theta - r_0 )^2 + \frac{\tilde{r}}{r_0} (r sin \theta)^2 \right] } r dr \; d \theta \\
E_{out}(\tilde{r}) &=& \frac{1}{i \lambda Z} \int_{\mathcal{C}_{(0,0)}^R} e^{i \frac{\pi}{\lambda Z} \left [ \frac{r_0}{\tilde{r} A(\tilde{r})^2} (x- r_0 )^2 + \frac{\tilde{r}}{r_0} y^2 \right] } dx \; dy
\end{eqnarray}
The main insight of this expansion is the fact that the propagation between the two mirrors of such a PIAA system reduces to the integration over an equivalent elliptical aperture. The geometry of this ellipse varies with $\tilde{r}$,  the location on $M2$. Qualitatively, based on an energy conservation argument, we already know that the local effective aperture size for propagation purposes is stretched by a factor of $\frac{\tilde{r} A(\tilde{r})^2}{r_0}$ in the radial direction. The elliptical integral in Eq.~\ref{EqEllipseNonCentNonStreched} formally illustrates this intuitive result, which states that since the area of integration has to be $A(\tilde{r})^2$, the effective local aperture size in the tangential direction, normal to the radial, has to shrink by a factor of $\frac{r_0}{\tilde{r}}$.
\section{Sensitivity to Off-Axis response}
\label{Sec:AppendixB}
The off-axis magnification of PIAA systems is the feature that makes such designs so appealing to the exo-planet community since it is  the source of their intrinsic extremely high angular resolution. It was first  explained by Guyon using energy and area conservation arguments (\cite{2003A&A...404..379G}) and then formally derived by Traub and Vanderbei (\cite{2003ApJ...599..695T}) using ray optics. Here we evaluate this magnification using Eq.~\ref{EqThPropAbber} in order to compute the sensitivity to off-axis sources.
Consider an off-axis source illuminating a PIAA system with a wavefront tilted by an angle $\gamma_{Sky}$. Assume that the resulting field distribution at $M2$ is Fourier transformed by an ideal lens. Call $\gamma_{Camera}$ the angular location of the centroid of the point spread function of such a source traveling through the pupil mapping unit.  Note that here we define the angular magnification using the centroid of the off-axis PSF and not its maximum. While this definition is less accurate it has the advantage of providing a number that does not vary with the angular separation of the off-axis source. We thus define $\mathcal{M}$, the angular magnification of a PIAA unit, as:
\begin{equation}
\mathcal{M} = \frac{\gamma_{Camera}}{\gamma_{Sky}}
\end{equation}
Then, given a two mirrors PIAA design, $\mathcal{M}$ can be computed as:
\begin{equation}
\mathcal{M} = \frac{1}{\pi R^2} \int_0^{2 \pi} \int_0^{R} A(\tilde{r})^2 \sqrt{\frac{r_0}{\tilde{r} A(\tilde{r})} \cos^2 \tilde{\theta} + \frac{\tilde{r}}{r_0} \sin^2\tilde{\theta}}  \; \tilde{r}d\tilde{r} d\theta 
\end{equation}
This result is a direct consequence of the ray optics remapping factor of Eq.~\ref{EqThPropAbber}. Without loss of generality we can assume that $\phi = 0$. If we write the angular separation of the off-axis source with respect to the optical axis in units of $\lambda /D$, then the terms corresponding to the geometric remapping in Eq.~\ref{EqThPropAbber} are written as:
\begin{equation}
E_{\alpha}(\tilde{r},\tilde{\theta}) = A(\tilde{r}) e^{i \frac{2 \pi}{D} \gamma_{Sky} (r_0 \cos \tilde{\theta})} = A(\tilde{r}) e^{i \frac{2 \pi}{D} \gamma_{Sky} x_0(\tilde{x},\tilde{y}) }
\end{equation}
For this calculation we leave out the equivalent angular spectrum factor since it will only change the phase of the companion. From Eq.~\ref{EqChainBis} we know  that:
\begin{eqnarray}
\frac{\partial x_0}{\partial \tilde{x}} &=& \frac{r_0}{\tilde{r} A(\tilde{r})} \cos^2\: \tilde{\theta} + \frac{\tilde{r}}{r_0} \sin^2\: \tilde{\theta} \\
\frac{\partial x_0}{\partial \tilde{y}} &=&  (\frac{r_0}{\tilde{r} A(\tilde{r})}  - \frac{\tilde{r}}{r_0}) cos\: \tilde{\theta} \sin\: \tilde{\theta} 
\end{eqnarray}
 Thus, at a given point on the surface of $M2$, $(\tilde{r},\tilde{\theta})$, the local spatial frequency is given by the magnitude of the gradient of $ \gamma_{Sky} \;  x_0(\tilde{x},\tilde{y})$
 \begin{eqnarray}
  N(\tilde{r},\tilde{\theta})& =& \gamma_{Sky} \sqrt{(\frac{\partial x_0}{\partial \tilde{x}})^2+(\frac{\partial x_0}{\partial \tilde{y}})^2} \nonumber \\
 &=& \gamma_{Sky} \sqrt{\frac{r_0}{\tilde{r} A(\tilde{r})} \cos^2 \tilde{\theta} + \frac{\tilde{r}}{r_0} \sin^2\tilde{\theta}}
 \label{EqLocalSFOff}
 \end{eqnarray}
 %
%
In order to compute $\mathcal{M}$ formally, we are interested in finding the centroid of this extended PSF.  More formally, the centroid of the planet PSF will be located at the barycenter of $ N(\tilde{r},\tilde{\theta})$ weighted by $A(\tilde{r})^2$, which gives:
\begin{equation}
\gamma_{Camera} = \langle N(\tilde{r},\tilde{\theta}) \rangle = \gamma_{Sky} \frac{1}{\pi R^2} \int_0^{2 \pi} \int_0^{R} A(\tilde{r})^2 \sqrt{\frac{r_0}{\tilde{r} A(\tilde{r})} cos^2 \tilde{\theta} + \frac{\tilde{r}}{r_0} sin^2\tilde{\theta}} \; \tilde{r} d\tilde{r} d\theta 
\end{equation}
%

When we use a PIAA unit that follows the $10^{-10}$ prolate profile for $A(\tilde{r})$, this yields $\mathcal{M} = 2.63$. This is exactly the value found when measuring the angular magnification using simulations such as the one shown on Fig.~\ref{figOffAxisPSFN24}, where we have computed the PSF of two off-axis sources, of respective angular separation $2$ and $4 \; \lambda/ D$, propagated through a PIAA unit. The centroids of these PSFs appear in the final image plane at $2 \times 2.63$ and $4 \times 2.63 \; \lambda/D$.  Note that because the main contribution to this angular magnification comes from the center of  $M2$, a legitimate approximation for this value is $\mathcal{M} \simeq A(0)$. A similar proof was presented by Guyon (\cite{2005ApJ...622..744G}), assuming ray optics. This result is a fundamental property of PIAA systems and is the source of their high performance. Because of the full throughput, spatial frequencies are magnified and planets that are very close to their parent star can be observed.

\section*{Acknowledgements}
The research described in this publication was carried out at the Jet Propulsion Laboratory, California Institute of Technology, under a contract with the National Aeronautics and Space Administration. The first author was supported by an appointment to the NASA Postdoctoral Program at the JPL, Caltech, administered by Oak Ridge Associated Universities through a contract with NASA. This work was also performed in part under contract with the California Institute of Technology (Caltech) funded by NASA through the Sagan Fellowship Program

\newpage

\section*{Captions}

\newpage

\subsection*{Figure 1}
Setup of the problem and notations]{Setup of the problem and notations. Top Left: Three dimensional representation of a pupil to pupil off-axis PIAA system. Top Right: Side view of the geometrical remapping in a pupil to pupil off-axis PIAA system. Bottom Left: Side view of the geometrical remapping in a pupil to pupil on-axis PIAA unit: \textbf{This is the configuration that is studied in this communication}. Bottom right: Side view of all the rays contributing to the diffractive  field at a point of coordinates $(\tilde{x},\tilde{y})$ at M2. The ray corresponding to the geometrical remapping, which has coordinates $(x_0(\tilde{x},\tilde{y}), y_0(\tilde{x},\tilde{y}))$ in the input plane, is highlighted.

\newpage

\subsection*{Figure 2}

Phase induced amplitude errors through a PIAA system. Spatial frequency at M1 $N=6$, with  $D = 3$ cm and $z = 1$

\newpage

\subsection*{Figure 3}

Phase induced amplitude errors through a PIAA system. Spatial frequency at M1 $N=10$, with  $\lambda = 700$ nm $D = 3$ cm and $z = 1$

\newpage

\subsection*{Figure 4}

Amplitude induced phase errors through a PIAA system. Spatial frequency at M1 $N=6$, with $\lambda = 700$ nm $D = 3$ cm and $z = 1$

\newpage

\subsection*{Figure 5}

Amplitude induced phase errors through a PIAA system. Spatial frequency at M1 $N=10$, with $\lambda = 700$ nm  $D = 3$ cm and $z = 1$

\newpage

\subsection*{Figure 6}

Cartoon representation of the wavelength data cube at $ M2$ -left panel-, and at the image plane -Right Panel. The transverse axis is a virtual cut across a wavelength cube. These cubes are obtained by stacking field distributions at $M2$ and the image plane for several wavelengths across the spectral bandwidth of interest.

\newpage

\subsection*{Figure 7}

Top: Illustration of the fit through the wavelength cube for one pixel. Top Left: raw contrast at one pixel in the image plane. Top Right: contrast at one pixel in the image plane after a perfect two sequential DM wavefront correction. Bottom: Residual intensity in the dark zone after subtracting the two dominant terms of the wavelength expansion. Note that the PSF of a ripple propagated through PIAA is much more extended than in the case of a classical coronagraph. The chromaticity of the leakage close to the optical axis has been modified by the propagator that introduced a higher order wavelength dependence. This drives the best speckle extinction achievable over a broadband. $N = 7$, $D = 3$ cm, $z = 1$ m, $\Delta \lambda / \lambda = 0.1$ 

\newpage

\subsection*{Figure 8}

Maximum of the broadband halo \textbf{in the dark hole} created by  two sequential DM wavefront controller  as a function of Fresnel number and spatial frequency of the wavefront error. The top curve corresponds to the maximum of the non-corrected PSF in the dark hole: note that high spatial frequencies leak in the dark hole due to the spatial extent of the off-axis PIAA PSF. The other three curves show the maximum of the residual halo after correction for, from top to bottom, $\mathcal{F}= 140, 1250, 11250$.

\newpage

\subsection*{Figure 9}

Maximum of the broadband halo \textbf{in the dark hole} created by  two sequential DM wavefront controller  as a function of  bandwidth and spatial frequency of the wavefront error. The top curve corresponds to the maximum of the non-corrected PSF in the dark hole: note that high spatial frequencies leak in the dark hole due to the spatial extent of the off-axis PIAA PSF. The other three curves show the maximum of the residual halo after correction for, from bottom to top, $\Delta \lambda / \lambda = 0.1, \; 0.2, \; 0.3$. The Fresnel number for the PIAA unit is $\mathcal{F}= 11250$.

\newpage

\subsection*{Figure 10}

Maximum of the broadband halo \textbf{in the dark hole} created by  two sequential DM wavefront controller  as a function of bandwidth and spatial frequency of the wavefront error. The top curve corresponds to the maximum of the non-corrected PSF in the dark hole: note that high spatial frequencies leak in the dark hole due to the spatial extent of the off-axis PIAA PSF. The other three curves show the maximum of the residual halo after correction for, from bottom to top, $\Delta \lambda / \lambda = 0.1, \; 0.2, \; 0.3$. The Fresnel number for the PIAA unit is $1250$

\newpage

\subsection*{Figure 11}

PSF of two off axis sources that are separated by $2 \lambda/ D$ and $4 \lambda/ D$ from the star

\newpage

\section*{Figures}

\newpage

\begin{figure}[h]
\includegraphics[width=6in]{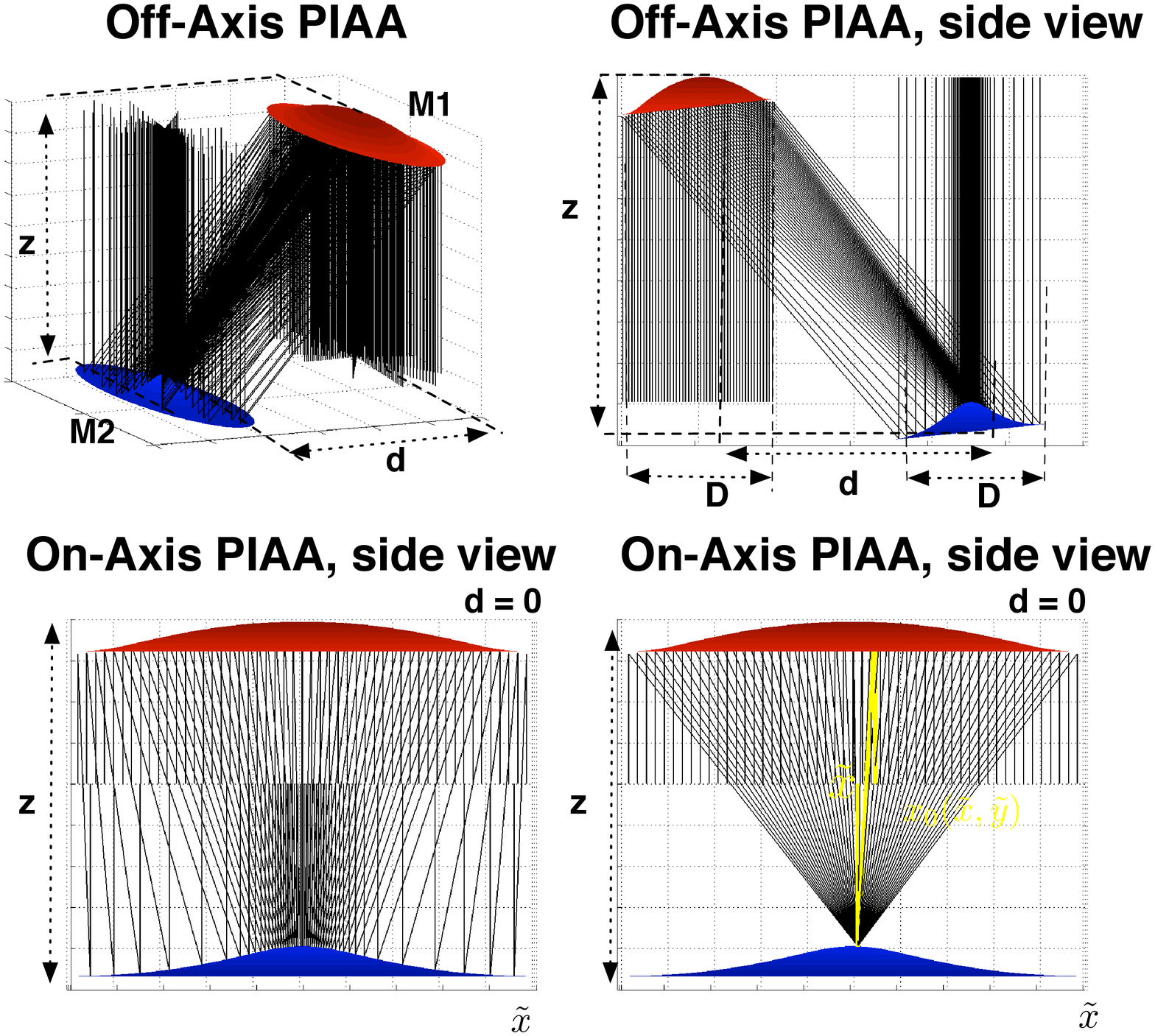}
\caption[Setup of the problem and notations]{Setup of the problem and notations. Top Left: Three dimensional representation of a
pupil to pupil off-axis PIAA system. Top Right: Side view of the geometrical remapping in a pupil to pupil off-axis PIAA system. Bottom Left: Side view of the geometrical remapping
in a pupil to pupil on-axis PIAA unit: \textbf{This is the configuration that is studied in this communication}. Bottom right: Side view of all the rays contributing
to the diffractive  field at a point of coordinates $(\tilde{x},\tilde{y})$ at M2. The ray corresponding to
the geometrical remapping, which has coordinates $(x_0(\tilde{x},\tilde{y}), y_0(\tilde{x},\tilde{y}))$ in the input plane, is
highlighted.}
\label{FigSetupPIAA}
\end{figure}

\newpage

\begin{figure}[h]
\begin{center}
    \includegraphics[width = 6in]{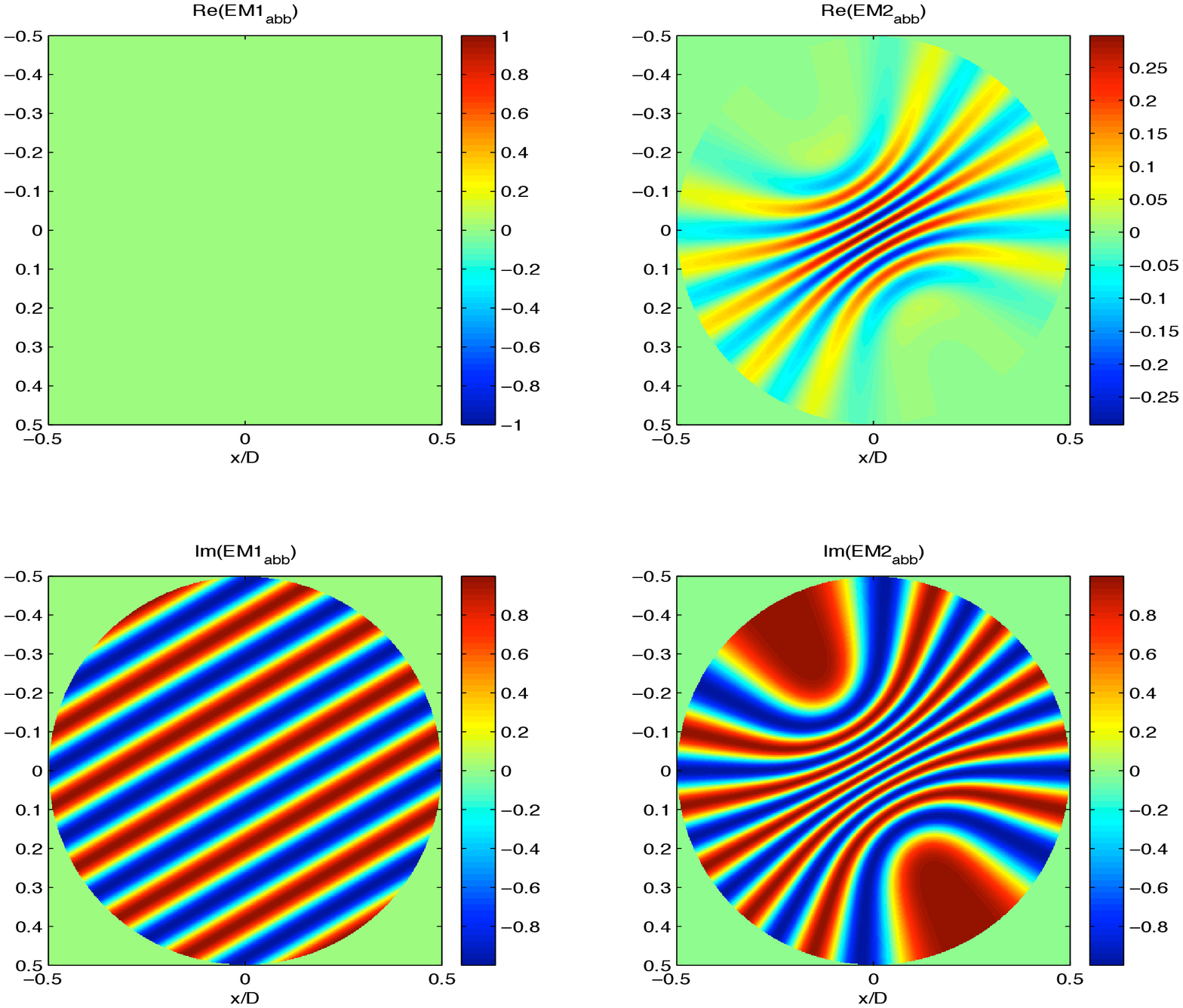}
\end{center}
\caption{Phase induced amplitude errors through a PIAA system. Spatial frequency at M1 $N=6$, with  $\lambda = 700$ nm  $D = 3$ cm and $z = 1$}
\label{figPhaseInducedAmplitudeN6}
\end{figure}

\newpage

\begin{figure}[h]
\begin{center}
    \includegraphics[width = 6in]{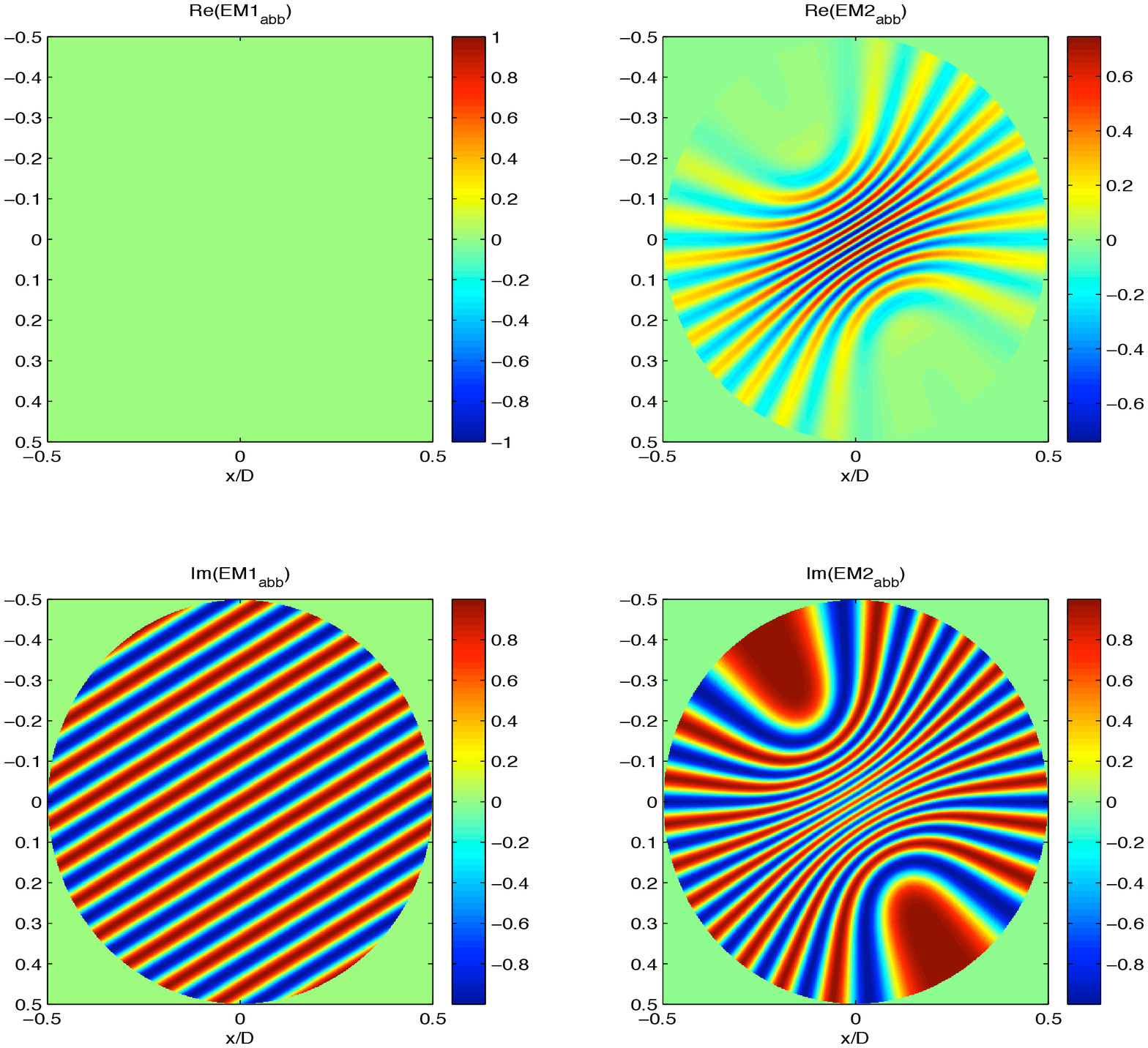}
\end{center}
\caption{Phase induced amplitude errors through a PIAA system. Spatial frequency at M1 $N=10$, with  $\lambda = 700$ nm $D = 3$ cm and $z = 1$}
\label{figPhaseInducedAmplitudeN10}
\end{figure}

\newpage

\begin{figure}[h]
\begin{center}
    \includegraphics[width = 6in]{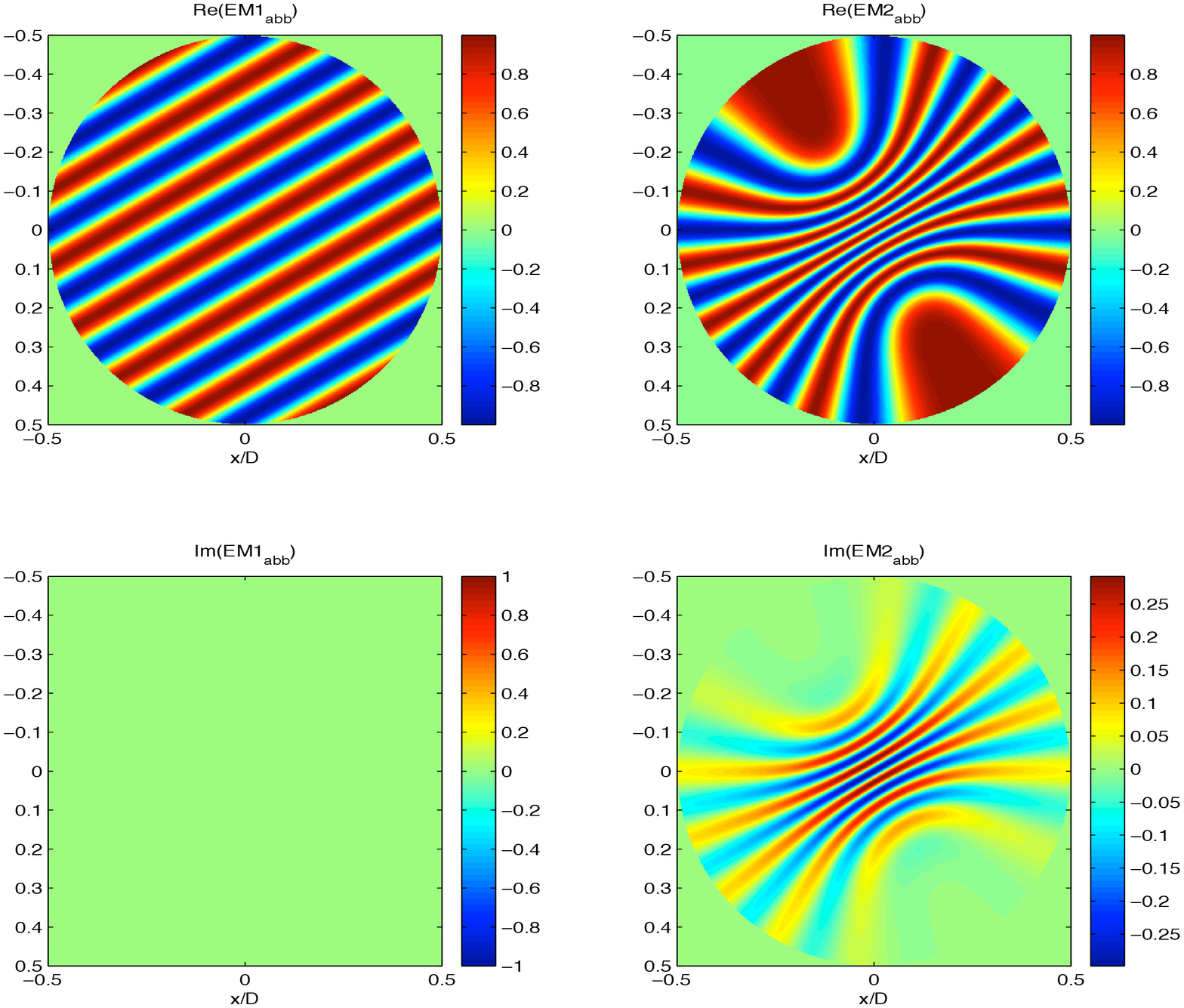}
\end{center}
\caption{Amplitude induced phase errors through a PIAA system. Spatial frequency at M1 $N=6$, with $\lambda = 700$ nm $D = 3$ cm and $z = 1$}
\label{figAmplitudeInducedPhaseN6}
\end{figure}

\newpage

\begin{figure}[h]
\begin{center}
    \includegraphics[width = 6in]{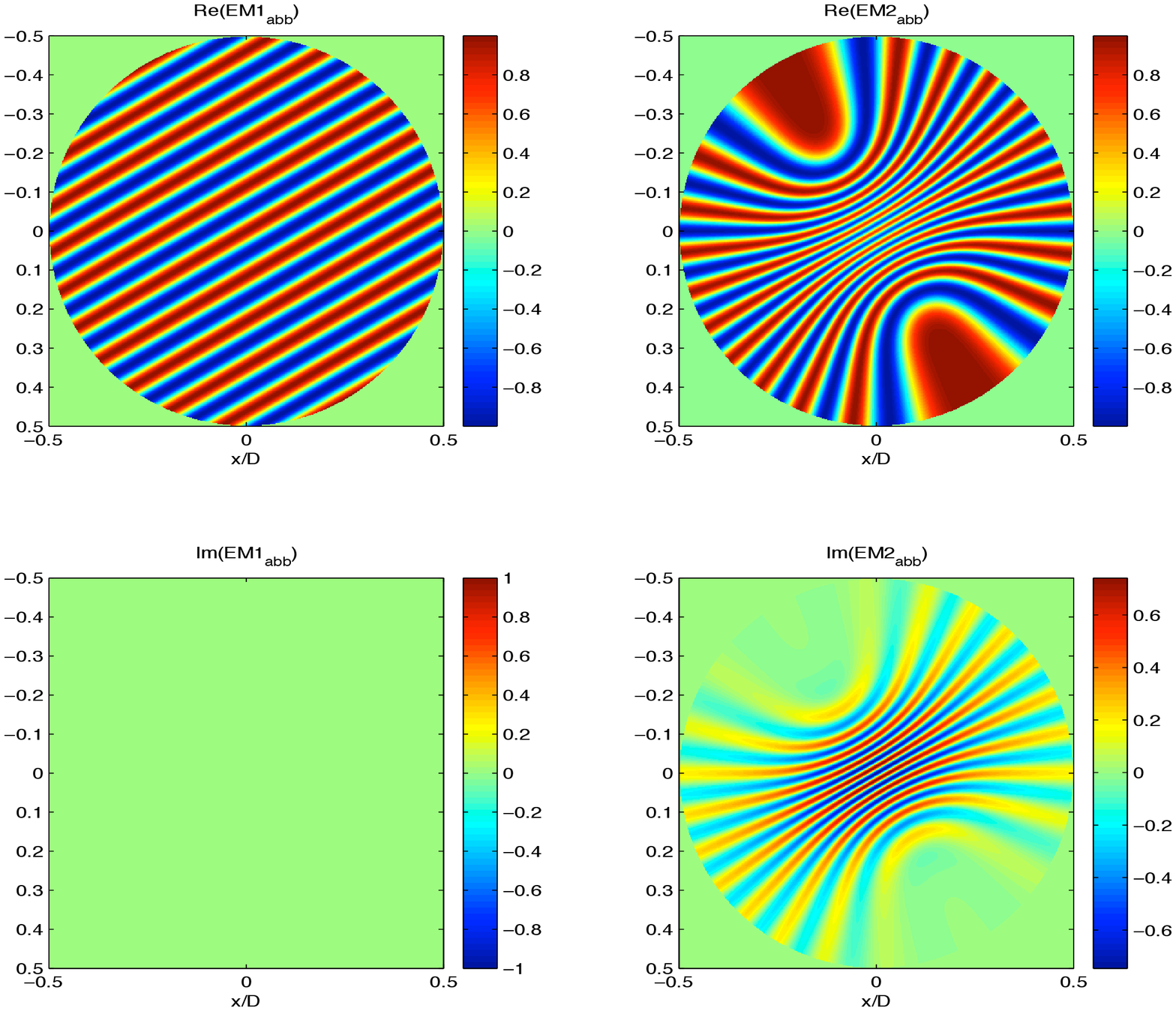}
\end{center}
\caption{Amplitude induced phase errors through a PIAA system. Spatial frequency at M1 $N=10$, with $\lambda = 700$ nm  $D = 3$ cm and $z = 1$}
\label{figAmplitudeInducedPhaseN10}
\end{figure}

\newpage

\begin{figure}[h]
\begin{center}
    \includegraphics[width = 6in]{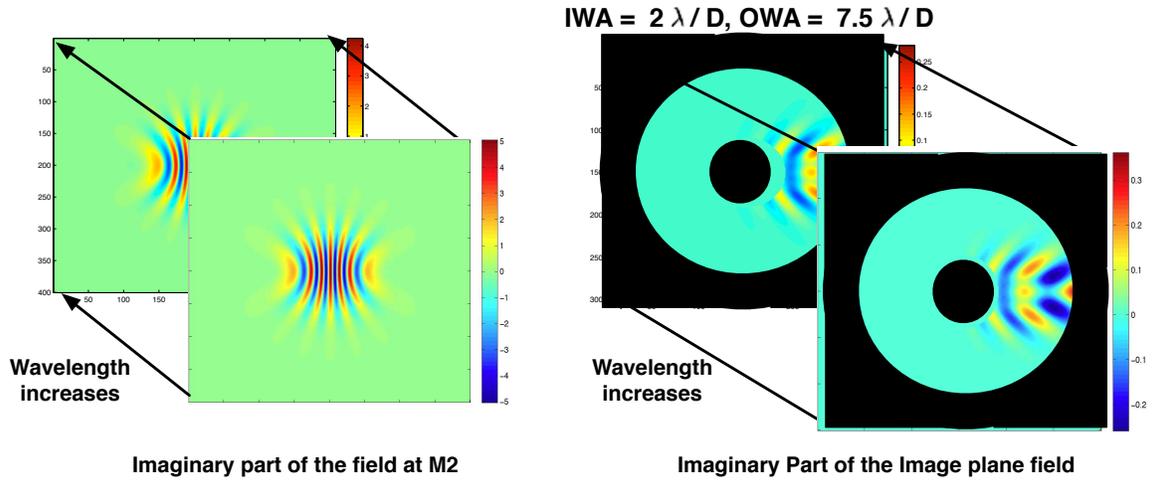}
\end{center}
\caption{Cartoon representation of the wavelength data cube at $ M2$ -left panel-, and at the image plane -Right Panel. The transverse axis is a virtual cut across a wavelength cube. These cubes are obtained by stacking field distributions at $M2$ and the image plane for several wavelengths across the spectral bandwidth of interest.}
\label{figPupilCubes}
\end{figure}

\newpage

\begin{figure}[h]
\begin{center}
     \includegraphics[width = 7in]{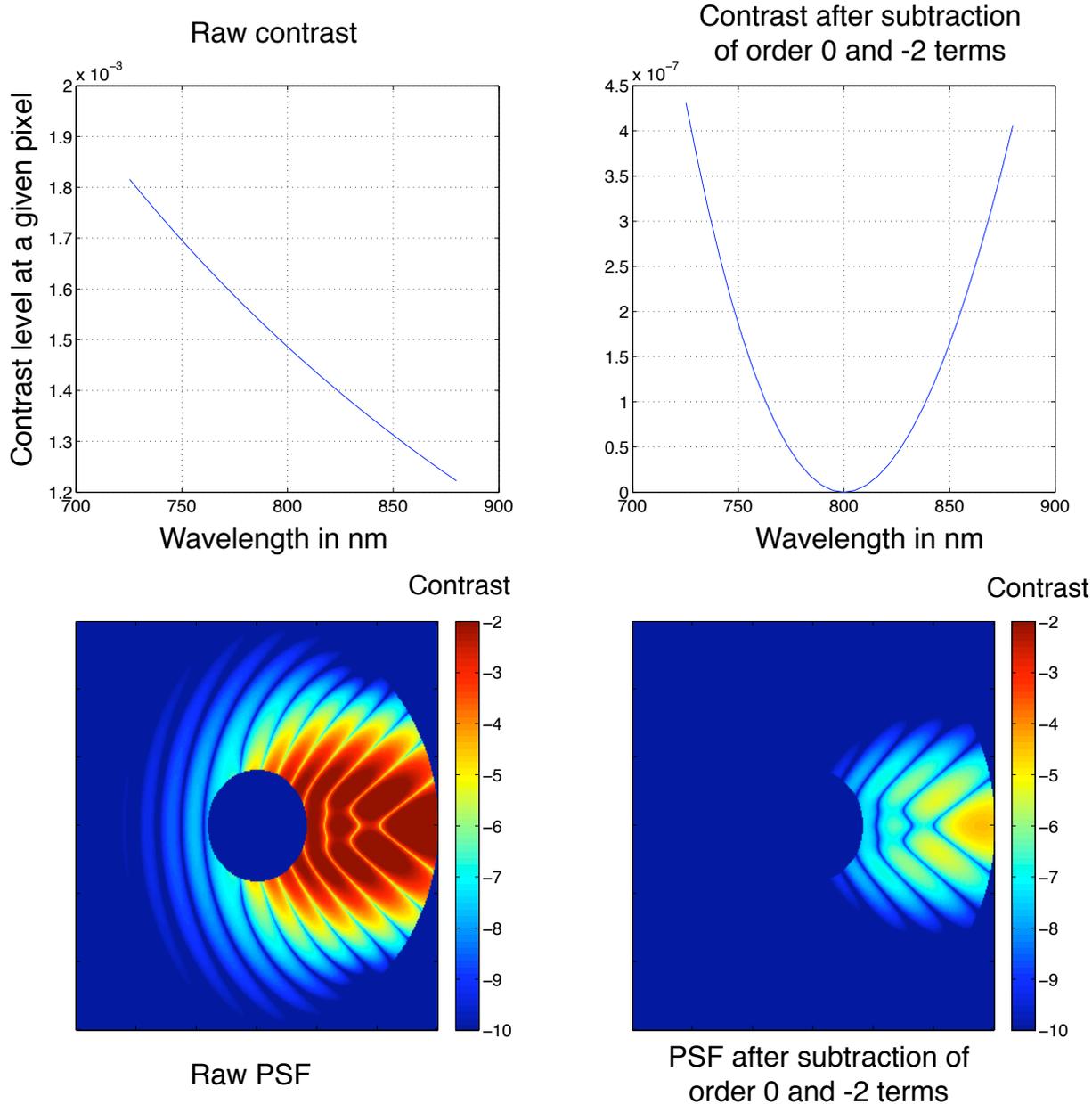}
\end{center}
\caption{Top: Illustration of the fit through the wavelength cube for one pixel. Top Left: raw contrast at one pixel in the image plane. Top Right: contrast at one pixel in the image plane after a perfect two sequential DM wavefront correction. Bottom: Residual intensity in the dark zone after subtracting the two dominant terms of the wavelength expansion. Note that the PSF of a ripple propagated through PIAA is much more extended than in the case of a classical coronagraph. The chromaticity of the leakage close to the optical axis has been modified by the propagator that introduced a higher order wavelength dependence. This drives the best speckle extinction achievable over a broadband. $N = 7$, $D = 3$ cm, $z = 1$ m, $\Delta \lambda / \lambda = 0.1$ }
\label{LambdFitTwoOrders}
\end{figure}

\newpage

\begin{figure}[h]
\begin{center}
    \includegraphics[width = 5in]{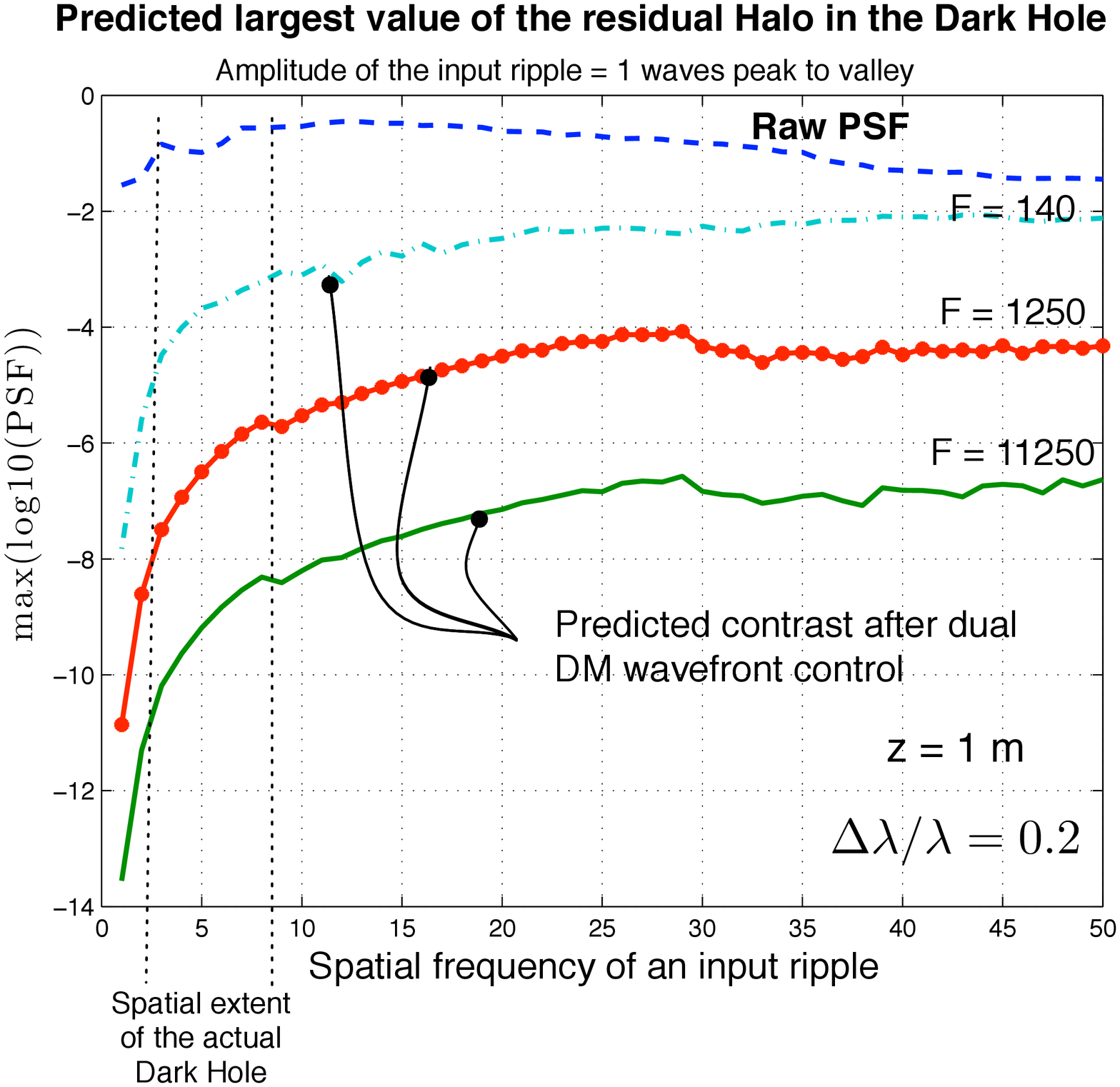}
\end{center}
\caption{Maximum of the broadband halo \textbf{in the dark hole} created by  two sequential DM wavefront controller  as a function of Fresnel number and spatial frequency of the wavefront error. The top curve corresponds to the maximum of the non-corrected PSF in the dark hole: note that high spatial frequencies leak in the dark hole due to the spatial extent of the off-axis PIAA PSF. The other three curves show the maximum of the residual halo after correction for, from top to bottom, $\mathcal{F}= 140, 1250, 11250$.}
\label{ContVSSfVSPupSize}
\end{figure}

\newpage

\begin{figure}[h]
\begin{center}
    \includegraphics[width = 5in]{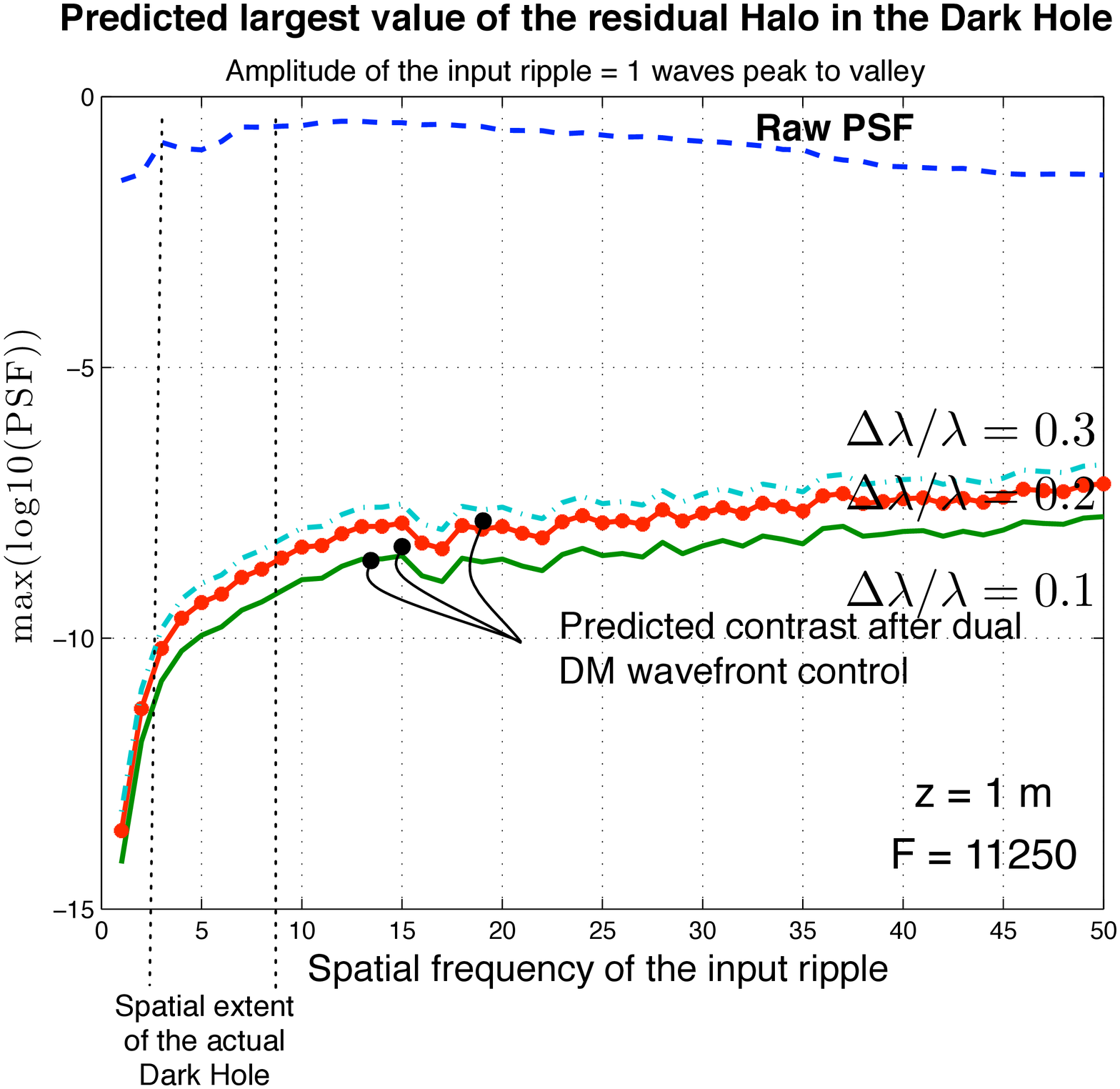}
\end{center}
\caption{Maximum of the broadband halo \textbf{in the dark hole} created by  two sequential DM wavefront controller  as a function of  bandwidth and spatial frequency of the wavefront error. The top curve corresponds to the maximum of the non-corrected PSF in the dark hole: note that high spatial frequencies leak in the dark hole due to the spatial extent of the off-axis PIAA PSF. The other three curves show the maximum of the residual halo after correction for, from bottom to top, $\Delta \lambda / \lambda = 0.1, \; 0.2, \; 0.3$. The Fresnel number for the PIAA unit is $\mathcal{F}= 11250$}
\label{ContVSSfVSBandwidth}
\end{figure}

\newpage

\begin{figure}[h]
\begin{center}
    \includegraphics[width = 5in]{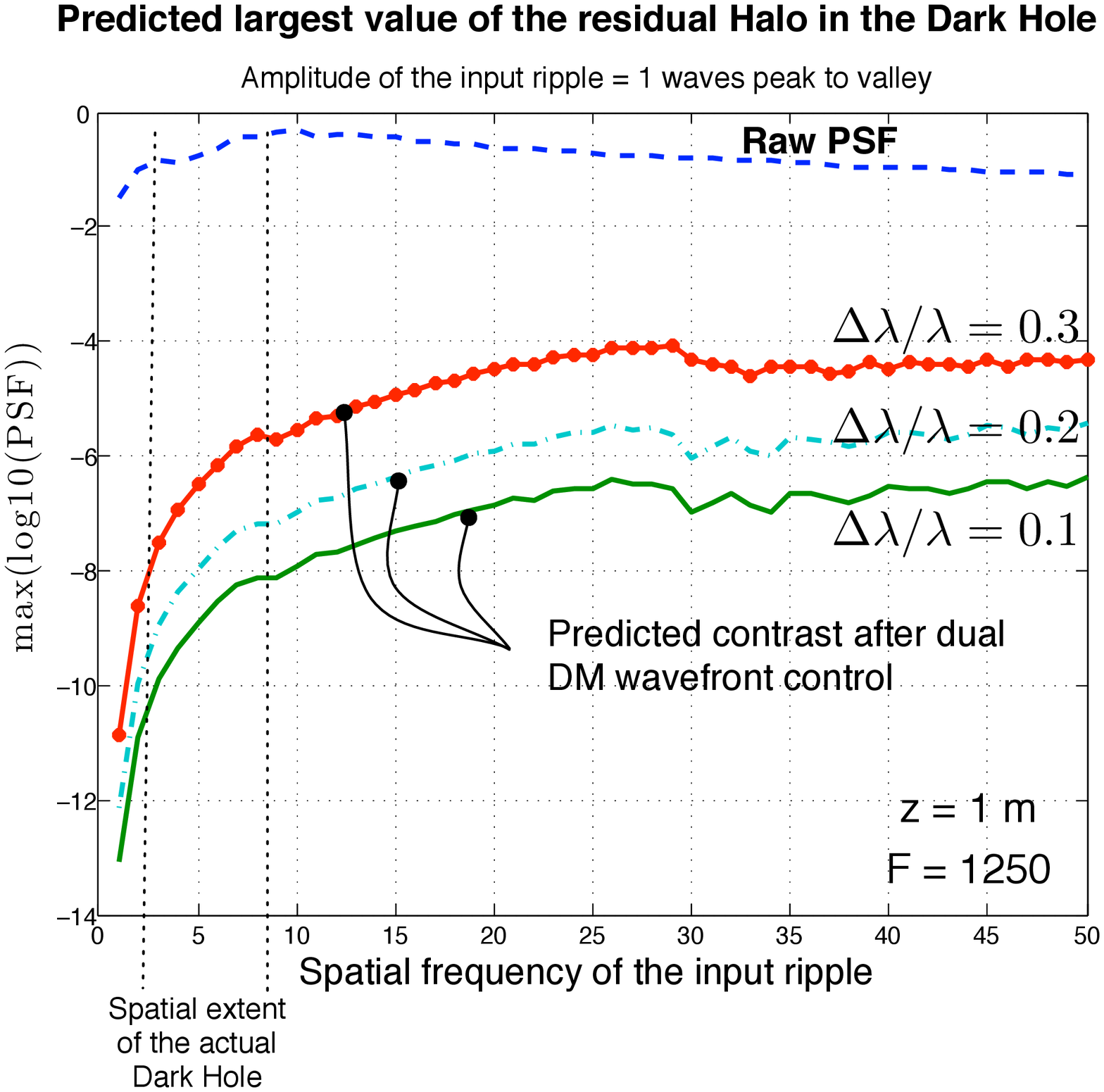}
\end{center}
\caption{Maximum of the broadband halo \textbf{in the dark hole} created by  two sequential DM wavefront controller  as a function of bandwidth and spatial frequency of the wavefront error. The top curve corresponds to the maximum of the non-corrected PSF in the dark hole: note that high spatial frequencies leak in the dark hole due to the spatial extent of the off-axis PIAA PSF. The other three curves show the maximum of the residual halo after correction for, from bottom to top, $\Delta \lambda / \lambda = 0.1, \; 0.2, \; 0.3$. The Fresnel number for the PIAA unit is $1250$}
\label{ContVSSfVSBandwidth32Legend}
\end{figure}

\newpage

\begin{figure}[h]
\begin{center}
    \includegraphics[width = 6in]{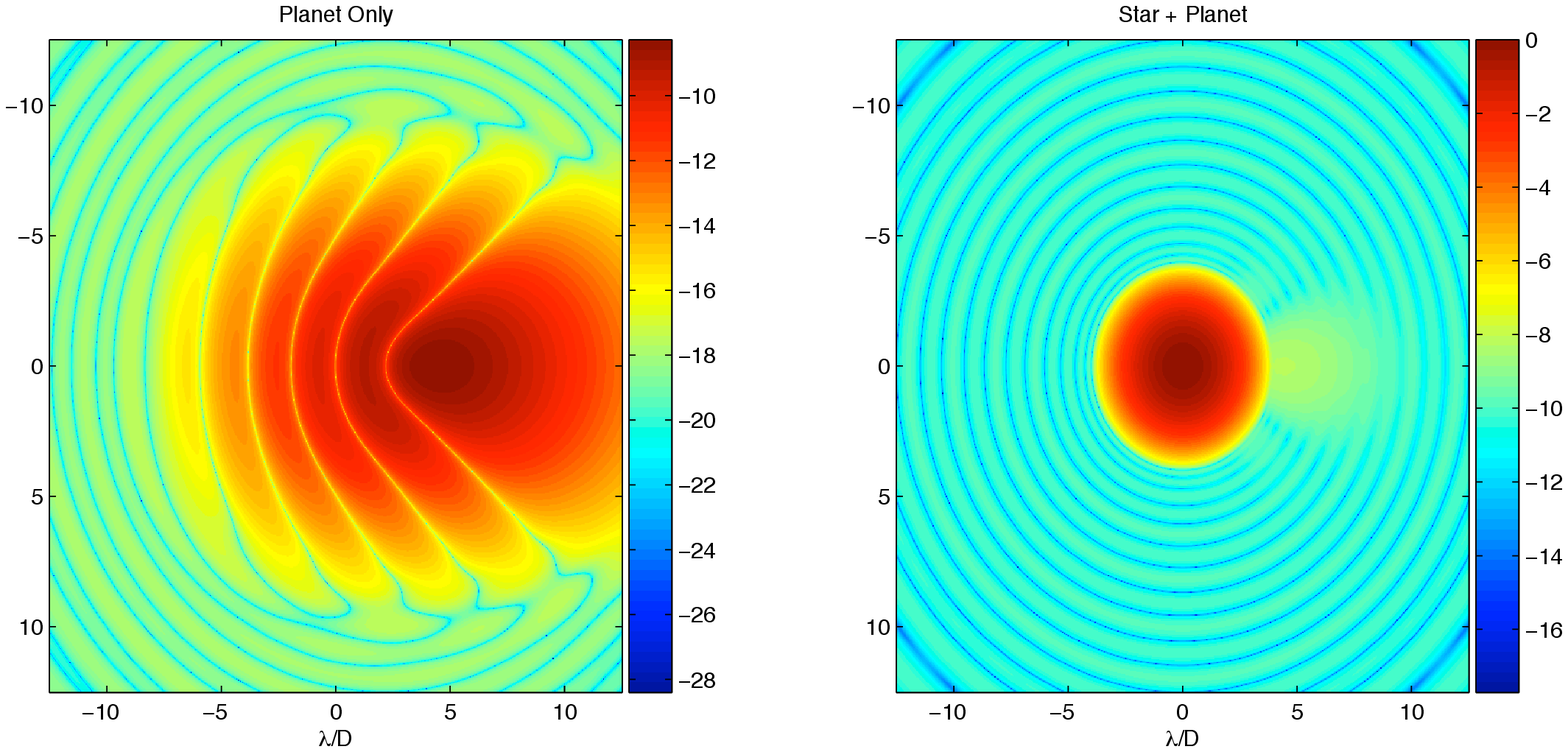}
    \includegraphics[width = 6in]{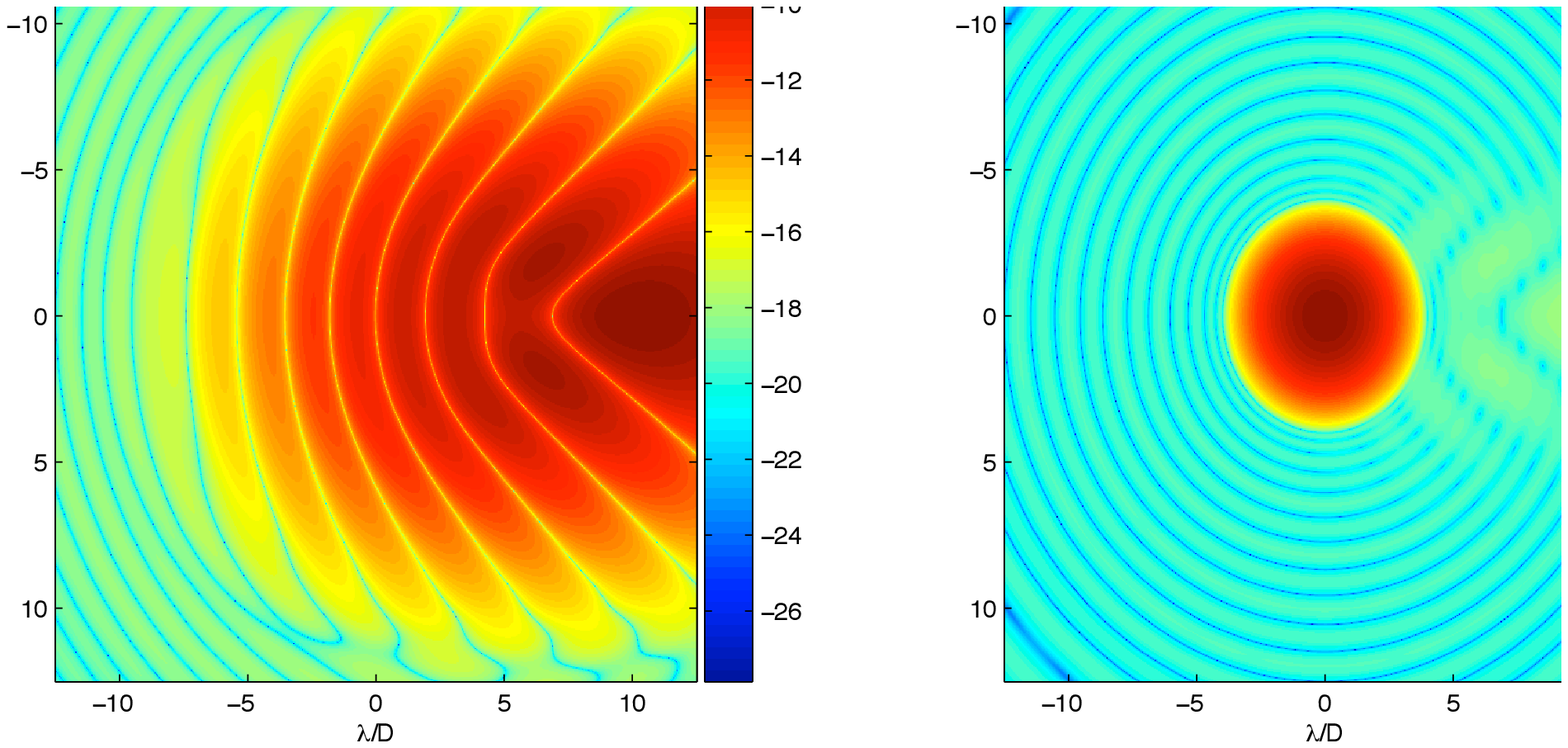}
\end{center}
\caption{PSF of two off axis sources that are separated by $2 \lambda/ D$ and $4 \lambda/ D$ from the star.}
\label{figOffAxisPSFN24}
\end{figure}

\end{document}